\documentclass[aps,prd,twocolumn,superscriptaddress]{revtex4}
\usepackage{epsfig,epsf}
\usepackage{amsmath}
\usepackage{amsthm}
\usepackage{amsfonts}
\usepackage{amssymb}
\usepackage{dsfont}
\usepackage{multirow}
\usepackage{appendix}
\usepackage{slashed}
\usepackage[active]{srcltx}
\usepackage{psfrag}

\begin{document}

\title{Exploring fully heavy scalar tetraquarks $QQ\overline{Q}\overline{Q}$}
\date{\today}
\author{S.~S.~Agaev}
\affiliation{Institute for Physical Problems, Baku State University, Az--1148 Baku,
Azerbaijan}
\author{K.~Azizi}
\affiliation{Department of Physics, University of Tehran, North Karegar Avenue, Tehran
14395-547, Iran}
\affiliation{Department of Physics, Do\v{g}u\c{s} University, Dudullu-\"{U}mraniye, 34775
Istanbul, T\"{u}rkiye}
\author{B.~Barsbay}
\affiliation{Division of Optometry, School of Medical Services and Techniques, Do\v{g}u%
\c{s} University, 34775 Istanbul, T\"{u}rkiye}
\author{H.~Sundu}
\affiliation{Department of Physics Engineering, Istanbul Medeniyet University, 34700
Istanbul, T\"{u}rkiye}

\begin{abstract}
The masses, current couplings and widths of the fully heavy scalar
tetraquarks $X_{\mathrm{4Q}}=QQ\overline{Q}\overline{Q}$, $Q=c, b$ are
calculated by modeling them as four-quark systems composed of axial-vector
diquark and antidiquark. The masses $m^{(\prime)}$ and couplings $%
f^{(\prime)}$ of these tetraquarks are computed in the context of the QCD
sum rule method by taking into account a nonperturbative term proportional
to the gluon condensate $\langle \alpha _{s}G^{2}/ \pi \rangle$. Results $%
m=(6570 \pm 55)~\mathrm{MeV}$ and $m^{\prime}=(18540 \pm 50)~\mathrm{MeV}$
are used to fix kinematically allowed hidden-flavor decay channels of these
states. It turns out that, the processes $X_{\mathrm{4c}}\rightarrow J/\psi
J/\psi $, $X_{\mathrm{4c}}\rightarrow \eta _{c}\eta _{c}$, and $X_{\mathrm{4c%
}}\rightarrow \eta _{c}\chi _{c1}(1P)$ are possible decay modes of $X_{%
\mathrm{4c}}$. The partial widths of these channels are evaluated by means
of the couplings $g_{i}, i=1,2,3$ which describe strong interactions of
tetraquark $X_{\mathrm{4c}}$ and mesons at relevant vertices. The couplings $%
g_{i}$ are extracted from the QCD three-point sum rules by extrapolating
corresponding form factors $g_{i}(Q^2) $ to the mass-shell of a final meson.
The mass of the scalar tetraquark $X_{\mathrm{4b}}$ is below the $\eta_b
\eta_b$ and $\Upsilon(1S)\Upsilon(1S)$ thresholds, therefore it does not
fall apart to these bottomonia, but transforms to conventional particles
through other mechanisms. Comparing $m=(6570 \pm 55)~\mathrm{MeV}$ and $%
\Gamma _{\mathrm{4c}}=(110 \pm 21)~\mathrm{MeV}$ with parameters of
structures observed by the LHCb, ATLAS and CMS collaborations, we interpret $%
X_{4c}$ as the resonance $X(6600)$ reported by CMS. Comparisons are made
with other theoretical predictions.
\end{abstract}

\maketitle


\section{Introduction}

\label{sec:Int} 

Conventional hadron spectroscopy encompasses variety of quark-antiquark
mesons and three-quark (antiquark) baryons with different contents and
spin-parities. But existence of multiquark particles composed of more than
three valence partons is not forbidden by any physical theory or model.
Features of such exotic states became object of theoretical studies just
after invention of quark-parton model and non-abelian field theory of strong
interactions.

Quantitative investigations of multiquark hadrons started from analyses
performed by Jaffe in Refs.\ \cite{Jaffe:1976ig,Jaffe:1976yi} using MIT
quark-bag model. In Ref.\ \cite{Jaffe:1976ig} he made an assumption about
four-quark $q^{2}\overline{q}^{2}$ nature of light mesons from the lowest
scalar nonet to explain the mass hierarchy of these particles. Another
intriguing result is connected with a state composed of six light quarks $%
S=uuddss$ \cite{Jaffe:1976yi}. This double-strange multiquark compound would
be stable against strong decays provided such particle really exists. Then
hexaquark $S$ may transform to ordinary hadrons only through weak processes
and, as a result, have mean lifetime $\tau \approx 10^{-10}\mathrm{s}$,
which is considerably longer than that of conventional mesons.

Stability against strong and/or electromagnetic decays is an important
question of exotic mesons's physics: Stable four-quark particles
(tetraquarks) with long mean lifetime may be discovered in various hadronic
processes relatively easily. Therefore, theoretical investigations of such
tetraquarks were and remain on agenda of high energy physics. Compounds
containing heavy $QQ$ diquarks ($Q=c$ or $b$ ) and light antidiquarks are
real candidates to stable exotic mesons. A group of hypothetical particles $%
QQ\overline{Q}^{(\prime )}\overline{Q}^{(\prime )}$ and $QQ\overline{q}%
\overline{q}$ were explored already in Refs.\ \cite%
{Ader:1981db,Lipkin:1986dw,Zouzou:1986qh}, in which it was shown that exotic
mesons built of only heavy quarks are unstable particles. But states with
content $QQ\overline{q}\overline{q}$ may form stable structures if the ratio
$m_{Q}/m_{q}$ is large. Conclusions about stable nature of the isoscalar
axial-vector tetraquark $T_{bb;\overline{u}\overline{d}}^{-}$ was also made
in Ref.\ \cite{Carlson:1987hh}, whereas four-quark mesons with heavy
diquarks $bc$ and $cc$ may be either stable or unstable particles.

More detailed analysis of fully heavy four-quark mesons $X_{\mathrm{4c}}=cc%
\overline{c}\overline{c}$, $X_{\mathrm{2bc}}=bc\overline{b}\overline{c}$ and
$X_{\mathrm{4b}}=bb\overline{b}\overline{b}$ was performed in Refs. \cite%
{Berezhnoy:2011xn,Karliner:2016zzc,Wu:2016vtq,Chen:2016jxd,Wang:2017jtz,Richard:2017vry}%
, in which different features of these particles were explored by means of
numerous methods and schemes. For instance, in Ref.\ \cite{Berezhnoy:2011xn}
masses of fully heavy tetraquarks were found by solving nonrelativistic
Schrodinger equation. In accordance with this article scalar and
axial-vector tetraquarks $X_{\mathrm{4c}}$, $X_{\mathrm{2bc}}$ are under the
di-$J/\psi $ and $J/\psi \Upsilon (1S)$ thresholds, and only tensor
particles can be seen in di-$J/\psi $ and $J/\psi \Upsilon (1S)$ invariant
mass distributions. At the same time, all fully beauty exotic mesons $X_{%
\mathrm{4b}}$ reside below $\Upsilon (1S)\Upsilon (1S)$ threshold, and
cannot be observed in this mass distribution. Masses of scalar tetraquarks $%
X_{\mathrm{4c}}$ and $X_{\mathrm{4b}}$ were estimated also in Ref.\ \cite%
{Karliner:2016zzc}. Results obtained there $m(X_{\mathrm{4c}})=(6192\pm 25)~%
\mathrm{MeV}$ and $m(X_{\mathrm{4b}})=(18826\pm 25)~\mathrm{MeV}$ allowed
the authors to study decay channels and productions of these particles.
Because $m(X_{\mathrm{4c}})$ is below di-$J/\psi $ but above $\eta _{c}\eta
_{c}$ thresholds, $X_{\mathrm{4c}}$ does not decay to $J/\psi J/\psi $
mesons, while a process $X_{\mathrm{4c}}\rightarrow \eta _{c}\eta _{c}$ is
the kinematically allowed mode. Similarly, $X_{\mathrm{4b}}$ cannot decay to
a pair of mesons $\Upsilon (1S)\Upsilon (1S)$, whereas $X_{\mathrm{4b}%
}\rightarrow \eta _{b}\eta _{b}$ is its possible channel. Interesting
predictions about particles $X_{\mathrm{4c}}$ and $X_{\mathrm{4b}}$ were
made in Ref. \cite{Chen:2016jxd}, in which masses of states $cc\overline{c}%
\overline{c}$, and $bb\overline{b}\overline{b}$ with different spin-parities
were calculated by applying the sum rule method. It was demonstrated that
masses of the scalar $J^{\mathrm{PC}}=0^{++}$ tetraquarks $X_{\mathrm{4c}}$
and $X_{\mathrm{4b}}$, except ones composed of pseudoscalar components, vary
inside limits $6.44-6.59\ \mathrm{GeV}$ and $18.45-18.59\ \mathrm{GeV}$,
respectively. Subsequently, $X_{\mathrm{4c}}$ decays to $\eta _{c}\eta _{c}$%
, $J/\psi J/\psi $, and $\eta _{c}\chi _{c1}(1P)$ meson pairs, whereas $X_{%
\mathrm{4b}}$ is stable against strong decays to hidden-beauty mesons:
Presumably a scalar diquark-antidiquark state $X_{\mathrm{4b}}$ built of
pseudoscalar components can decay to $\eta _{b}\eta _{b}$ and $\Upsilon
(1S)\Upsilon (1S)$ mesons. In accordance with Ref. \cite{Wang:2017jtz}, the
scalar and tensor $X_{\mathrm{4c}}$, have masses $5.99~\mathrm{GeV}$ and $%
6.09~\mathrm{GeV}$, and decay to mesons $\eta _{c}\eta _{c}$, whereas di-$%
J/\psi $ channel is forbidden for them.

Experimental studies of two charmonia or bottomonia productions in $pp$ and $%
p\overline{p}$ collisions provided valuable information on nature and decay
channels of fully heavy exotic mesons. Thus, a pair of $J/\psi $ mesons were
observed by LHCb, CMS and D0 Collaborations \cite%
{LHCb:2011kri,CMS:2014cmt,D0:2014vql}, respectively. The $J/\psi \Upsilon
(1S)$ and $\Upsilon (1S)\Upsilon (1S)$ pairs were detected and investigated
by D0 and CMS experiments \cite{D0:2015dyx,CMS:2016liw}. In the four-quark
picture such final states imply production of intermediate states $cc%
\overline{c}\overline{c}$, $bc\overline{b}\overline{c}$ and $bb\overline{b}%
\overline{b}$ with their subsequent decays to couples of heavy conventional
mesons.

The discovery of the doubly charmed baryon $\Xi _{cc}^{++}=ccu$ by the LHCb
Collaboration \cite{Aaij:2017ueg} gave strong impetus to investigations of
doubly and fully heavy tetraquarks. Thus, the mass of $\Xi _{cc}^{++}$ was
used as an input parameter to estimate the mass of the axial-vector
tetraquark $T_{bb;\overline{u}\overline{d}}^{-}$ \cite{Karliner:2017qjm}.
Conclusions about strong-interaction stable nature of the tetraquarks $bb%
\overline{u}\overline{d}$, $bb\overline{u}\overline{s}$, and $bb\overline{d}%
\overline{s}$ were made on the basis of heavy-quark symmetry as well \cite%
{Eichten:2017ffp}. Weak decays of stable double-heavy tetraquarks were
explored in numerous publications \cite%
{Xing:2018bqt,Li:2018bkh,Agaev:2018khe,Agaev:2020mqq,Agaev:2020dba,Agaev:2019kkz,Sundu:2019feu, Agaev:2019lwh,Agaev:2020zag,Yu:2017pmn}%
. In our articles \cite%
{Agaev:2018khe,Agaev:2020mqq,Agaev:2020dba,Agaev:2019kkz,Sundu:2019feu,
Agaev:2019lwh,Agaev:2020zag}, we calculated masses and current couplings of
the tetraquarks $bb\overline{u}\overline{d}$, $bb\overline{u}\overline{s}$
and $bc\overline{u}\overline{d}$ with spin-parities $J^{\mathrm{P}}=0^{+}$,$%
\ 1^{+}$, as well as parameters of the scalar state $bs\overline{u}\overline{%
d} $. We evaluated full width of these structures by considering their
numerous semileptonic and nonleptonic weak decay channels.

The class of fully heavy exotic mesons $QQ\overline{Q}^{(\prime )}\overline{Q%
}^{(\prime )}$ were also explored in Refs.\ \cite%
{Hughes:2017xie,Esposito:2018cwh,Anwar:2017toa,Bai:2016int,Liu:2019zuc}.
Predictions some of these papers \cite{Anwar:2017toa,Bai:2016int} confirm in
a modified form the results discussed above. But there are also publications
which contradict to such conclusions. In fact, using lattice simulations the
authors of Ref.\ \cite{Hughes:2017xie} did not find evidence for tetraquarks
$X_{\mathrm{4b}}$ with different spin-parities below the lowest thresholds
in relevant channels.

Recently, LHCb reported new structures in the di-$J/\psi $ mass distribution
extracted from $pp$ data at c.m. energies $7$, $8$, and $13~\mathrm{TeV}$
\cite{LHCb:2020bwg}. The LHCb observed a threshold enhancement in
nonresonant di-$J/\psi $ production from $6.2$ to $6.8~\mathrm{GeV}$ with
center at $6.49~\mathrm{GeV}$. A narrow peak at $6.9~\mathrm{GeV}$, and a
resonance around $7.2~\mathrm{GeV}$ were seen as well. The narrow state
labeled $X(6900)$ has parameters
\begin{eqnarray}
m_{1}^{\mathrm{LHCb}} &=&(6905\pm 11\pm 7)~\mathrm{MeV,}  \notag \\
\Gamma _{1}^{\mathrm{LHCb}} &=&(80\pm 19\pm 33)~\mathrm{MeV,}  \label{eq:MW1}
\end{eqnarray}%
when assuming no interference with nonresonant single-parton scattering
(NRSPS) continuum, and
\begin{eqnarray}
m_{2}^{\mathrm{LHCb}} &=&(6886\pm 11\pm 11)~\mathrm{MeV,}  \notag \\
\Gamma _{2}^{\mathrm{LHCb}} &=&(168\pm 33\pm 69)~\mathrm{MeV,}
\label{eq:MW2}
\end{eqnarray}%
while ones takes into account interference of NRSPS with a threshold
enhancement.

This experimental information was detailed and extended by the ATLAS and CMS
Collaborations \cite{Bouhova-Thacker:2022vnt,CMS:2023owd}. The ATLAS
announced three resonances $X(6200)$, $X(6600)$, and $X(6900)$ in the di-$%
J/\psi $ channel with the parameters%
\begin{eqnarray}
m_{0}^{\mathrm{ATL}} &=&6220\pm 50_{-50}^{+40}~\mathrm{MeV,}  \notag \\
\Gamma _{0}^{\mathrm{ATL}} &=&310\pm 120_{-80}^{+70}~\mathrm{MeV,}
\label{eqMWATL1}
\end{eqnarray}%
\begin{eqnarray}
m_{1}^{\mathrm{ATL}} &=&6620\pm 30_{-10}^{+20}~\mathrm{MeV,}  \notag \\
\Gamma _{1}^{\mathrm{ATL}} &=&310\pm 90_{-110}^{+60}~\mathrm{MeV,}
\label{eq:MWATL2}
\end{eqnarray}%
and%
\begin{eqnarray}
m_{2}^{\mathrm{ATL}} &=&6870\pm 30_{-10}^{+60}~\mathrm{MeV,}  \notag \\
\Gamma _{2}^{\mathrm{ATL}} &=&120\pm 40_{-10}^{+30}~\mathrm{MeV.}
\label{eq:MWATL3}
\end{eqnarray}%
The resonance $X(7300)$ with the mass and width%
\begin{eqnarray}
m_{3}^{\mathrm{ATL}} &=&7220\pm 30_{-30}^{+20}~\mathrm{MeV,}  \notag \\
\Gamma _{3}^{\mathrm{ATL}} &=&100_{-70-50}^{+130+60}~\mathrm{MeV,}
\end{eqnarray}%
was fixed in the $J/\psi \psi ^{\prime }$ channel. The resonances $X(6200)$
and $X(6600)$ belong to an enhancement in a $6.2-6.8~\mathrm{GeV}$ region
observed by LHCb. It seems reasonable to suppose that LHCb fixed a
superposition of these structures. The resonance $X(7300)$ is close to
structure at $7.2~\mathrm{GeV}$ reported by LHCb.

Resonances $X(6600)$, $X(6900)$ and $X(7300)$ discovered by CMS and analyzed
in the no-interference model have the following masses and widths
\begin{eqnarray}
m_{1}^{\mathrm{CMS}} &=&(6552\pm 10\pm 12)~\mathrm{MeV},  \notag \\
\Gamma _{1}^{\mathrm{CMS}} &=&(124_{-26}^{+32}\pm 33)~\mathrm{MeV},
\label{eq:MWCMS1}
\end{eqnarray}%
\begin{eqnarray}
m_{2}^{\mathrm{CMS}} &=&(6927\pm 9\pm 4)~\mathrm{MeV},  \notag \\
\Gamma _{2}^{\mathrm{CMS}} &=&(122_{-21}^{+24}\pm 18)~\mathrm{MeV},
\label{eq:MWCMS2}
\end{eqnarray}%
and%
\begin{eqnarray}
m_{3}^{\mathrm{CMS}} &=&(7287_{-18}^{+20}\pm 5)~\mathrm{MeV},  \notag \\
\Gamma _{3}^{\mathrm{CMS}} &=&(95_{-40}^{+59}\pm 19)~\mathrm{MeV},
\label{eq:MWCMS3}
\end{eqnarray}%
respectively. Summing up, we can state that there are four resonances in the
range $6.2-7.3~\mathrm{GeV}$ discovered by different collaborations in the
di-$J/\psi $ and $J/\psi \psi ^{\prime }$ mass distributions.

Observations made by LHCb stimulated further detailed studies of fully heavy
exotic mesons \cite{Zhang:2020xtb,Wang:2020ols,Wang:2020dlo,
Albuquerque:2020hio,Yang:2020wkh,Becchi:2020mjz,Becchi:2020uvq,Dong:2020nwy,Dong:2021lkh,Liang:2021fzr}%
. Needless to say, that all models and technical tools available in high
energy physics were activated to explore these problems. Interesting results
concerning properties of fully heavy tetraquarks were obtained using the sum
rule method in Refs.\ \cite{Zhang:2020xtb,Wang:2020ols,Wang:2020dlo,
Albuquerque:2020hio,Yang:2020wkh}. For example, depending on a type of
interpolating current, the mass of the scalar tetraquark $cc\overline{c}%
\overline{c}$ was found within limits $6.44-6.47~\mathrm{GeV}$ \cite%
{Zhang:2020xtb}. Fully heavy diquark-antidiquark and hadronic molecules were
analyzed also in Ref.\ \cite{Albuquerque:2020hio}, in which the resonance $%
X(6900)$ was interpreted as a molecule $\chi _{c0}\chi _{c0}$ or/and a
tetraquark built of pseudoscalar ingredients.

The LHCb data were considered in Ref.\ \cite{Dong:2020nwy} in the framework
of a coupled-channel approach: It was argued that in the di-$J/\psi $ system
exists a near-threshold state $X(6200)$ with spin-parities $0^{++}$ or $%
2^{++}$. Coupled-channel effects may also generate a pole structure
identified in Ref.\ \cite{Liang:2021fzr} with the resonance $X(6900)$.
Analysis performed there allowed the authors also to predict existence of a
bound state $X(6200)$, and broad and narrow resonances $X(6680)$ and $%
X(7200) $, respectively.

Information of the ATLAS and CMS Collaborations considerably clarified
status of structures above the di-$J/\psi $ threshold, and generated new
interesting assumptions about their nature \cite%
{Wang:2022xja,Faustov:2022mvs,Niu:2022vqp,Dong:2022sef,Yu:2022lak,Kuang:2023vac}%
. Indeed, in Ref.\ \cite{Wang:2022xja} the $X(6200)$ was assigned to be the
ground-level tetraquark state with $J^{\mathrm{PC}}=0^{++}$ or $1^{+-}$,
whereas its first radial excitation was interpreted as $X(6600)$. Using the
relativized Godfrey-Isgur diquark model, the authors of Ref.\ \cite%
{Dong:2022sef} proposed to consider the resonances starting from $X(6200)$
as the $1S$, $1P/2S$, $1D/2P$, and $2D/3P/4S$ tetraquark states. Similar
interpretations were suggested in the context of the relativistic quark
model as well \cite{Faustov:2022mvs}.

As is seen, there are numerous alternatives to describe structures reported
by the different collaborations. In present article, we address problems of
these new data, and explore the fully charmed tetraquark $X_{\mathrm{4c}}$
with $J^{\mathrm{PC}}=0^{++}$ by calculating its mass, current coupling and
width. We model $X_{\mathrm{4c}}$ as a diquark-antidiquark structure, and
apply the two-point sum rule method to calculate a relevant correlation
function including a nonperturbative term $\sim \langle \alpha _{s}G^{2}/\pi
\rangle $. It turns out, that processes $X_{\mathrm{4c}}\rightarrow J/\psi
J/\psi $, $X_{\mathrm{4c}}\rightarrow \eta _{c}\eta _{c}$, and $X_{\mathrm{4c%
}}\rightarrow \eta _{c}\chi _{c1}(1P)$ are allowed decay modes of $X_{%
\mathrm{4c}}$. To calculate their partial widths, we make use of the
three-point sum rule approach, and compute strong form factors $%
g_{i}(q^{2}),\ i=1,2,3$ describing interaction of particles at vertices $X_{%
\mathrm{4c}}J/\psi J/\psi $, $X_{\mathrm{4c}}\eta _{c}\eta _{c}$, and $X_{%
\mathrm{4c}}\eta _{c}\chi _{c1}(1P)$, respectively. Predictions for strong
couplings $g_{i}$, obtained after extrapolation of $g_{i}(q^{2})$ to the
mass-shell of a final meson, are used to calculate widths of aforementioned
decay channels and to estimate full width $\Gamma _{\mathrm{4c}}$ of the
tetraquark $X_{\mathrm{4c}}$. Such detailed information places
interpretation of $X_{\mathrm{4c}}$ on strong bases and leads to reliable
conclusions. We evaluate also the mass $m^{\prime }$ of the state $X_{%
\mathrm{4b}}$ and show that in the axial--axial model $X_{\mathrm{4b}}$ is
stable against strong decays to two bottomonia. It is worth noting that in
the present paper we do not consider other mechanisms of $X_{\mathrm{4c}}$
and $X_{\mathrm{4b}}$ decays to conventional particles \cite%
{Becchi:2020mjz,Becchi:2020uvq}.

This article is structured in the following way: In Section \ref{sec:Masses}%
, we calculate masses and current couplings of the tetraquarks $X_{\mathrm{4c%
}}$ and $X_{\mathrm{4b}}$. Strong decay of $X_{\mathrm{4c}}$ to $J/\psi
J/\psi $ is considered in Sec.\ \ref{sec:Decays1}. Partial widths of the
processes $X_{\mathrm{4c}}\rightarrow \eta _{c}\eta _{c}$ and $X_{\mathrm{4c}%
}\rightarrow \eta _{c}\chi _{c1}(1P)$ are computed in Sec.\ \ref{sec:Decays2}%
. Here, we find also the full width $\Gamma _{\mathrm{4c}}$ of the
tetraquark $X_{\mathrm{4c}}$. Last section is reserved for discussion of
results and concluding notes. Appendix contains the explicit expression of
the heavy-quark propagator, and the perturbative part of the spectral
density used in mass computations.


\section{Spectroscopic parameters of the tetraquarks $X_{\mathrm{4c}}$ and $%
X_{\mathrm{4b}}$}

\label{sec:Masses}

In this section, we calculate the masses $m^{(\prime )}$ and current
couplings $f^{(\prime )}$ of the tetraquarks $X_{\mathrm{4c}}$ and $X_{%
\mathrm{4b}}$ by means of the QCD two-point sum rule approach \cite%
{Shifman:1978bx,Shifman:1978by}. It is a powerful nonperturbative method
developed to investigate features of conventional mesons and baryons, but
can also be applied to study multiquark hadrons, such as tetraquarks and
pentaquarks.

To derive the sum rules necessary for extracting the masses and current
couplings of the scalar tetraquarks $X_{\mathrm{4c}}$ and $X_{\mathrm{4b}}$,
we begin from analysis of the two-point correlation function
\begin{equation}
\Pi (p)=i\int d^{4}xe^{ipx}\langle 0|\mathcal{T}\{J(x)J^{\dag
}(0)\}|0\rangle .  \label{eq:CF1}
\end{equation}%
where, $\mathcal{T}$ \ is the time-ordered product of two currents, and $%
J(x) $ is the interpolating currents for these states.

We model the tetraquarks $X_{\mathrm{4c}}$ and $X_{\mathrm{4b}}$ as
structures formed by the axial-vector diquark $Q^{T}C\gamma _{\mu }Q$ and
axial-vector antidiquark $\overline{Q}\gamma _{\mu }C\overline{Q}^{T}$.
Corresponding interpolating current is given by the formula
\begin{equation}
J(x)=Q_{a}^{T}(x)C\gamma _{\mu }Q_{b}(x)\overline{Q}_{a}(x)\gamma ^{\mu }C%
\overline{Q}_{b}^{T}(x),  \label{eq:CR1}
\end{equation}%
where $a$, and $b$ are color indices. In Eq.\ (\ref{eq:CR1}) $Q(x)$ denotes
either $c$ or $b$ quark fields, and $C$ is the charge conjugation matrix.
The current $J(x)$ describes the tetraquark with spin-parities $J^{\mathrm{PC%
}}=0^{++}$.

In what follows, we write down formulas for the tetraquark $X_{\mathrm{4c}}$%
: Expressions for the state $X_{\mathrm{4b}}$ can be obtained from them
trivially. The physical side of the sum rule $\Pi ^{\mathrm{Phys}}(p)$
\begin{equation}
\Pi ^{\mathrm{Phys}}(p)=\frac{\langle 0|J|X_{\mathrm{4c}}(p)\rangle \langle
X_{\mathrm{4c}}(p)|J^{\dagger }|0\rangle }{m^{2}-p^{2}}+\cdots,
\label{eq:Phys1}
\end{equation}%
is derived from Eq.\ (\ref{eq:CF1}) by inserting a complete set of
intermediate states with quark content and spin-parities of the tetraquark $%
X_{\mathrm{4c}}$, and performing integration over $x$. Let us note that in $%
\Pi ^{\mathrm{Phys}}(p)$ the ground-state term is written down explicitly,
whereas contributions of higher resonances and continuum states are shown by
the dots.

The correlation function $\Pi ^{\mathrm{Phys}}(p)$ can be simplified using
the matrix element
\begin{equation}
\langle 0|J|X_{\mathrm{4c}}(p)\rangle =fm,  \label{eq:ME1}
\end{equation}%
which leads to the following expression
\begin{equation}
\Pi ^{\mathrm{Phys}}(p)=\frac{f^{2}m^{2}}{m^{2}-p^{2}}+\cdots .
\label{eq:Phen2}
\end{equation}%
The correlator $\Pi ^{\mathrm{Phys}}(p)$ has simple Lorentz structure
proportional to $\mathrm{I}$, therefore the invariant amplitude $\Pi ^{%
\mathrm{Phys}}(p^{2})$ is given by right-hand side of Eq.\ (\ref{eq:Phen2}).

The QCD side of the sum rule $\Pi ^{\mathrm{OPE}}(p)$ has to be computed in
the operator product expansion ($\mathrm{OPE}$) with certain accuracy. For
these purposes, one substitutes the current $J(x)$ into the correlator $\Pi
(p)$, contracts relevant quark fields, and replaces contractions by the
heavy quark propagators. These manipulations lead to the formula
\begin{eqnarray}
&&\Pi ^{\mathrm{OPE}}(p)=i\int d^{4}xe^{ipx}\left\{ \mathrm{Tr}\left[ \gamma
_{\mu }\widetilde{S}_{c}^{b^{\prime }b}(-x)\gamma _{\nu }S_{c}^{a^{\prime
}a}(-x)\right] \right.  \notag \\
&&\times \left[ \mathrm{Tr}\left[ \gamma ^{\nu }\widetilde{S}%
_{c}^{aa^{\prime }}(x)\gamma ^{\mu }S_{c}^{bb^{\prime }}(x)\right] -\mathrm{%
Tr}\left[ \gamma ^{\nu }\widetilde{S}_{c}^{ba^{\prime }}(x)\gamma ^{\mu
}\right. \right.  \notag \\
&&\left. \left. \times S_{c}^{ab^{\prime }}(x)\right] \right] +\mathrm{Tr}%
\left[ \gamma _{\mu }\widetilde{S}_{c}^{a^{\prime }b}(-x)\gamma _{\nu
}S_{c}^{b^{\prime }a}(-x)\right]  \notag \\
&&\left. \times \left[ \mathrm{Tr}\left[ \gamma ^{\nu }\widetilde{S}%
_{c}^{ba^{\prime }}(x)\gamma ^{\mu }S_{c}^{ab^{\prime }}(x)\right] -\mathrm{%
Tr}\left[ \gamma ^{\nu }\widetilde{S}_{c}^{aa^{\prime }}(x)\gamma ^{\mu
}S_{c}^{bb^{\prime }}(x)\right] \right] \right\},  \notag \\
&&  \label{eq:QCD1}
\end{eqnarray}%
where%
\begin{equation}
\widetilde{S}_{c}(x)=CS_{c}^{T}(x)C,  \label{eq:Prop}
\end{equation}%
with $S_{c}(x)$ being the $c$-quark propagator. The explicit expression of
the heavy quark propagator $S_{Q}(x)$ can be found in Appendix.

In the case under analysis, the QCD side of the sum rules depends
exclusively on the propagators of heavy quarks. The heavy quark propagator $%
S_{Q}^{ab}(x)$ apart from a perturbative term contains also components which
are linear and quadratic in the gluon field strength. It does not depend on
light quark or mixed quark-gluon vacuum condensates which are sources of
main nonperturbative contributions to correlation functions.

The $\Pi ^{\mathrm{OPE}}(p)$ has simple Lorentz structure $\sim \mathrm{I}$
as well. In what follows, the corresponding invariant amplitude will be
denoted by $\Pi ^{\mathrm{OPE}}(p^{2})$. Having equated two functions $\Pi ^{%
\mathrm{Phys}}(p^{2})$ and $\Pi ^{\mathrm{OPE}}(p^{2})$, applied the Borel
transformation to suppress contributions of higher resonances and continuum
states, and subtracted these contributions by employing the assumption about
quark-hadron duality \cite{Shifman:1978bx,Shifman:1978by}, we find the
required sum rules for the mass and coupling of the tetraquark $X_{\mathrm{4c%
}}$.

Calculation of the function $\Pi ^{\mathrm{OPE}}(p^{2})$ is a next step in
our efforts to derive the sum rules for $m$ and $f$. Analyses demonstrate
that after the Borel transformation and continuum subtraction the amplitude $%
\Pi (M^{2},s_{0})$ has the form%
\begin{equation}
\Pi (M^{2},s_{0})=\int_{16m_{c}^{2}}^{s_{0}}ds\rho ^{\mathrm{OPE}%
}(s)e^{-s/M^{2}}.  \label{eq:InvAmp}
\end{equation}%
Here, $\rho ^{\mathrm{OPE}}(s)$ is a two-point spectral density, which is
found as an imaginary part of the invariant amplitude $\Pi ^{\mathrm{OPE}%
}(p^{2})$. The function $\rho ^{\mathrm{OPE}}(s)$ contains a perturbative
term $\rho ^{\mathrm{pert.}}(s)$ and a dimension-$4$ nonperturbative
contribution proportional to $\langle \alpha _{s}G^{2}/\pi \rangle $. In
Appendix, we write down the analytical expression for $\rho ^{\mathrm{pert.}%
}(s)$, and refrain from presenting a dimension-$4$ term which is rather
lengthly.

Then, the sum rules for $m$ and $f$ \ are given by the formulas
\begin{equation}
m^{2}=\frac{\Pi ^{\prime }(M^{2},s_{0})}{\Pi (M^{2},s_{0})}  \label{eq:Mass}
\end{equation}%
and
\begin{equation}
f^{2}=\frac{e^{m^{2}/M^{2}}}{m^{2}}\Pi (M^{2},s_{0}),  \label{eq:Coupl}
\end{equation}%
respectively. In Eq.\ (\ref{eq:Mass}), we use the notation $\Pi ^{\prime
}(M^{2},s_{0})=d\Pi (M^{2},s_{0})/d(-1/M^{2})$.

The sum rules Eqs.\ (\ref{eq:Mass}) and (\ref{eq:Coupl}) depend on the gluon
vacuum condensate and on masses of $c$ and $b$ quarks, numerical values of
which are listed below%
\begin{eqnarray}
&&\langle \frac{\alpha _{s}G^{2}}{\pi }\rangle =(0.012\pm 0.004)~\mathrm{GeV}%
^{4},  \notag \\
&&\ m_{c}=(1.27\pm 0.02)~\mathrm{GeV},  \notag \\
&&m_{b}=4.18_{-0.02}^{+0.03}~\mathrm{GeV}.  \label{eq:Parameters}
\end{eqnarray}

Choosing working windows for parameters $M^{2}$ and $s_{0}$ is another
problem of the sum rule computations. They should be fixed in such a way
that to meet a constraint imposed on the pole contribution ($\mathrm{PC}$),
and ensure convergence of the operator product expansion. Because, in the
present article we consider only a nonperturbative term $\sim \langle \alpha
_{s}G^{2}/\pi \rangle $, the pole contribution plays a decisive role in
determining of $M^{2}$ and $s_{0}$. To estimate $\mathrm{PC}$, we use the
expression%
\begin{equation}
\mathrm{PC}=\frac{\Pi (M^{2},s_{0})}{\Pi (M^{2},\infty )},  \label{eq:PC}
\end{equation}%
and require fulfillment of the constraint $\mathrm{PC}\geq 0.5$.

\begin{figure}[h]
\includegraphics[width=8.5cm]{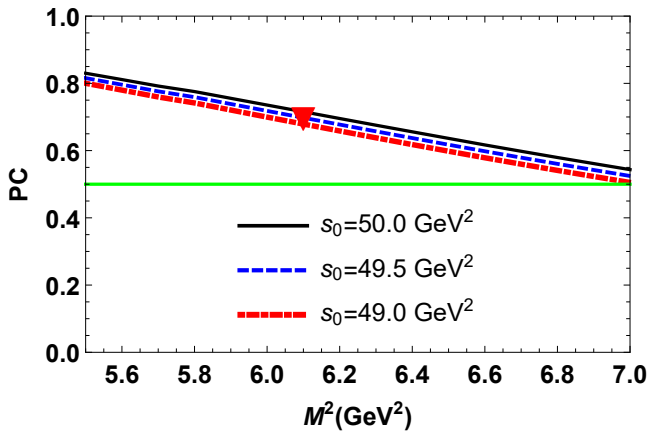}
\caption{The pole contribution $\mathrm{PC}$ as a function of the Borel
parameter $M^{2}$ at different $s_{0}$. The limit $\mathrm{PC}=0.5$ is shown
by the horizontal line. The red triangle shows the point, where the mass $m$
of the tetraquark $X_{\mathrm{4c}}$ has been extracted from the sum rule. }
\label{fig:PC}
\end{figure}

The $\mathrm{PC}$ is employed to fix the higher limit of the Borel parameter
$M^{2}$. The lower limit for $M^{2}$ is found from a stability of the sum
rules' results under variation of $M^{2}$, and from prevalence of the
perturbative term. Two values of $M^{2}$ extracted by this method fix
boundaries of the region where $M^{2}$ can be varied. Calculations for the
tetraquark $X_{\mathrm{4c}}$ show that the intervals
\begin{equation}
M^{2}\in \lbrack 5.5,7]~\mathrm{GeV}^{2},\ s_{0}\in \lbrack 49,50]~\mathrm{%
GeV}^{2},  \label{eq:Wind1}
\end{equation}%
are appropriate for the parameters $M^{2}$ and $s_{0}$, and comply with
limits on $\mathrm{PC}$ and nonperturbative term. Thus, at $M^{2}=7~\mathrm{%
GeV}^{2}$ the pole contribution is $0.51$, whereas at $M^{2}=5.5~\mathrm{GeV}%
^{2}$ it becomes equal to $0.82$. At the minimum of $M^{2}=5.5~\mathrm{GeV}%
^{2}$, contribution of the nonperturbative term is negative and forms $2\%$
of the correlation function. To demonstrate dynamics of the pole
contribution, Fig.\ \ref{fig:PC} we plot $\mathrm{PC}$ as a function of $%
M^{2}$ at different $s_{0}$. It is seen, that the pole contribution exceeds $%
0.5$ for all values of the parameters $M^{2}$ and $s_{0}$ from Eq.\ (\ref%
{eq:Wind1}).

\begin{widetext}

\begin{figure}[h!]
\begin{center}
\includegraphics[totalheight=6cm,width=8cm]{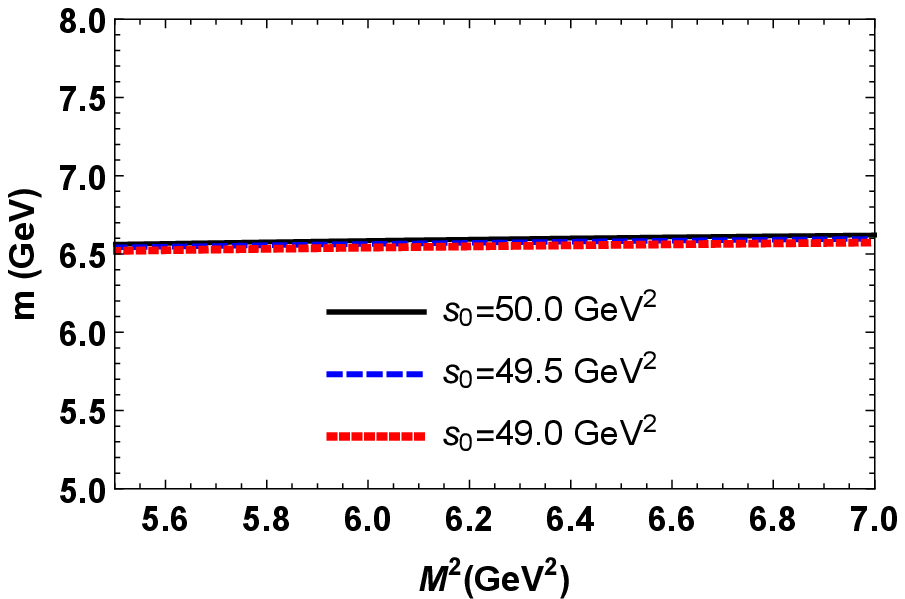}\,\, %
\includegraphics[totalheight=6cm,width=8cm]{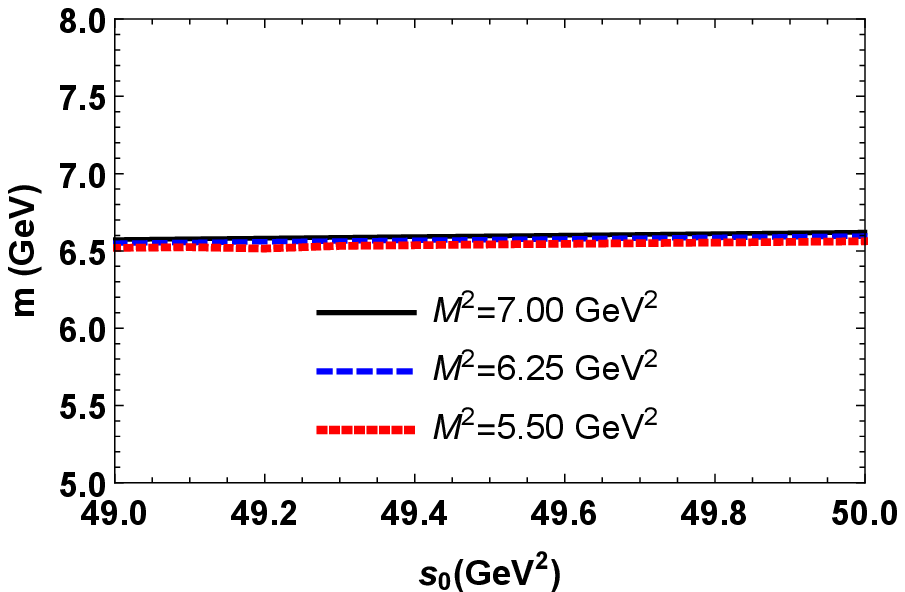}
\end{center}
\caption{ Mass of the tetraquark $X_{\mathrm{4c}}$ as a function of the
Borel parameter $M^2$ (left), and as a function
of the continuum threshold $s_0$ (right).}
\label{fig:Mass}
\end{figure}

\end{widetext}

We extract the mass $m$ and coupling $f$ of the tetraquark $X_{\mathrm{4c}}$
by calculating them at different $M^{2}$ and $s_{0}$, and determining their
mean values averaged over the regions Eq.\ (\ref{eq:Wind1}). Our predictions
for $m$ and $f$ read
\begin{eqnarray}
m &=&(6570\pm 55)~\mathrm{MeV},  \notag \\
f &=&(5.61\pm 0.39)\times 10^{-2}~\mathrm{GeV}^{4}.  \label{eq:Result1}
\end{eqnarray}%
The results in Eq.\ (\ref{eq:Result1}) correspond to sum rules predictions
at approximately middle point of the regions in Eq.\ (\ref{eq:Wind1}), i.e.,
to predictions at the point $M^{2}=6.1~\mathrm{GeV}^{2}$ and $s_{0}=49.5~%
\mathrm{GeV}^{2}$, where the pole contribution is $\mathrm{PC}\approx 0.70$.
This fact guarantees the dominance of $\mathrm{PC}$ in the obtained results,
and confirms ground-state nature of the tetraquark $X_{\mathrm{4c}}$.
Dependence of $m$ on the parameters $M^{2}$ and $s_{0}$ is depicted in Fig.\ %
\ref{fig:Mass}.

The mass $m$ of the tetraquark $X_{\mathrm{4c}}$ obtained in present article
nicely agrees with the mass of the resonance $X(6600)$ fixed by the ATLAS
and CMS collaborations, and belong to the wide threshold enhancement $%
6.2-6.8~\mathrm{GeV}$ in $J/\psi J/\psi $ mass distribution seen by LHCb.
Therefore, at this level of our knowledge, we consider the tetraquark $X_{%
\mathrm{4c}}$ as a candidate to the $X(6600)$ state. But, for more detailed
comparisons with ATLAS and CMS data, and credible statements about its
nature, we need to evaluate the full width of $X_{\mathrm{4c}}$.

\begin{widetext}

\begin{figure}[h!]
\begin{center}
\includegraphics[totalheight=6cm,width=8cm]{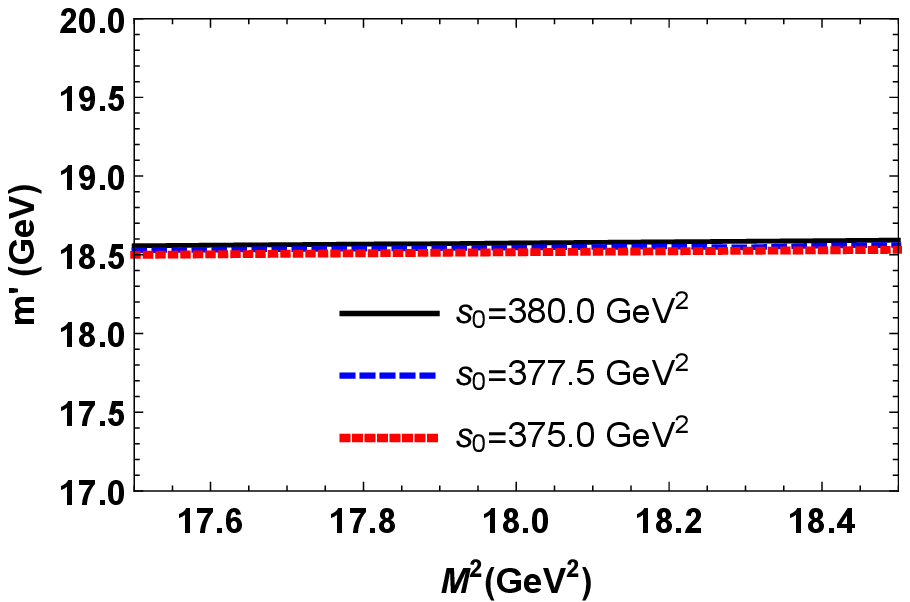}\,\, %
\includegraphics[totalheight=6cm,width=8cm]{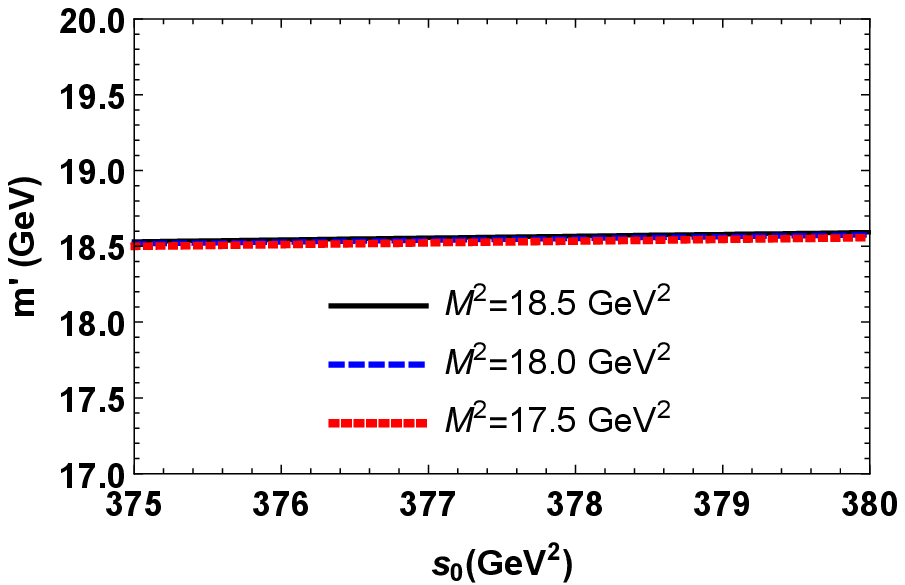}
\end{center}
\caption{ The same as in Fig.\ 2, but for the mass $m^{\prime}$ of the tetraquark $X_{\mathrm{4b}}$.}
\label{fig:MassB}
\end{figure}

\end{widetext}

In the case of the tetraquark $X_{\mathrm{4b}}$ a similar analysis yields
the following working intervals for the Borel and continuum subtraction
parameters
\begin{eqnarray}
M^{2} &\in &[17.5,18.5]~\mathrm{GeV}^{2},\   \notag \\
s_{0} &\in &[375,380]~\mathrm{GeV}^{2}.  \label{eq:Wind2}
\end{eqnarray}%
The pole contribution in the interval for $M^{2}$ changes within limits%
\begin{equation}
0.72\geq \mathrm{PC}\geq 0.66.
\end{equation}%
At $M^{2}=17.5~\mathrm{GeV}^{2}$ the dimension-$4$ term constitutes $\simeq
-1.5\%$ of the result. The mass and current coupling of $X_{\mathrm{4b}}$
are
\begin{eqnarray}
m^{\prime } &=&(18540\pm 50)~\mathrm{MeV},  \notag \\
f^{\prime } &=&(6.1\pm 0.4)\times 10^{-1}~\mathrm{GeV}^{4}.
\label{eq:Result2}
\end{eqnarray}%
Behavior of $m^{\prime }$ as a function of $M^{2}$ and $s_{0}$ is shown in
Fig.\ \ref{fig:MassB}.

Fully beauty scalar tetraquarks were investigated in numerous articles. The
mass $m^{\prime }$ of the scalar $4b$ state was found equal to $18754~%
\mathrm{MeV}$, $(18826\pm 25)~\mathrm{MeV,}$ and $18450-18590\ \mathrm{MeV}$
in Refs.\ \cite{Berezhnoy:2011xn,Karliner:2016zzc,Chen:2016jxd},
respectively. These results were obtained by solving nonrelativistic
Schrodinger equation, using a phenomenological approach or the QCD sum rule
method. An estimate $18750\ \mathrm{MeV}$ for the mass of the $4b$
ground-state particle was made in the context of a relativized diquark model
with one-gluon-exchange and confining potentials \cite{Anwar:2017toa}. A
diffusion Monte Carlo method used to solve nonrelativistic many-body system
led to the result $m^{\prime }=(18690\pm 30)~\mathrm{MeV}$ \cite{Bai:2016int}%
. Considerably larger mass $19315~\mathrm{MeV}$ was predicted for the $J^{%
\mathrm{PC}}=0^{++}$ diquark-antidiquark state in Ref.\ \cite%
{Faustov:2022mvs}.

These results differ from each other not only quantitatively, but imply also
different mechanisms for decays of these particles. Thus, there are two
important thresholds for fully beauty tetraquarks, i.e., the $2\eta _{b}$
and $2\Upsilon (1S)$ thresholds that amount to $18798~\mathrm{MeV}$ and $%
18921~\mathrm{MeV}$, respectively. Possible decay modes of $4b$ four-quark
compounds to ordinary mesons and leptons are determined by their positions
in this mass scale.

Our result $m^{\prime }=18540~\mathrm{MeV}$ for the mass of $X_{\mathrm{4b}}$
is consistent with prediction of Ref.\ \cite{Chen:2016jxd} calculated also
in the framework of the sum rule method. It is below the lowest $2\eta _{b}$
threshold in the sector of fully beauty ordinary mesons. In other words, $X_{%
\mathrm{4b}}$ is stable against strong decays to conventional $b\overline{b}$
mesons. Similar conclusions were drawn also in Refs. \cite%
{Berezhnoy:2011xn,Chen:2016jxd}. Such structures transform to conventional
particles due to $b\overline{b}$ annihilation to a gluon or a light
quark-antiquark pair, through two and three gluons produced by a $b\overline{%
b}$ pair which later are converted into light hadrons \cite{Becchi:2020mjz}.
In Ref.\ \cite{Becchi:2020mjz} the width of the fully beauty tetraquark with
the mass below the $2\eta _{b}$ threshold was estimated around of $8.5\
\mathrm{MeV}$. Hence, the tetraquark $X_{\mathrm{4b}}$ has a finite width
though it does not fall apart to $2\eta _{b}$ and $2\Upsilon (1S)$ final
states, but processes that generate this width are beyond the scope of the
present work.

\section{Decay $X_{\mathrm{4c}}\rightarrow J/\protect\psi J/\protect\psi $}

\label{sec:Decays1}


The mass of the tetraquark $X_{\mathrm{4c}}$ exceeds the two-meson
thresholds both in $J/\psi J/\psi $ and $\eta _{c}\eta _{c}$ channels,
therefore $S$-wave processes $X_{\mathrm{4c}}\rightarrow J/\psi J/\psi $ and
$X_{\mathrm{4c}}\rightarrow \eta _{c}\eta _{c}$ are allowed decay modes of
this particle. Another channel which will be considered in the present
article is $P$-wave decay mode $X_{\mathrm{4c}}\rightarrow \eta _{c}\chi
_{c1}(1P)$.

We begin our investigations from analysis of the process $X_{\mathrm{4c}%
}\rightarrow J/\psi J/\psi $. The partial width of this decay is determined
by the strong coupling $g_{1}$ of the particles at the vertex $X_{\mathrm{4c}%
}J/\psi J/\psi $. In the context of the QCD sum rule method $g_{1}$ can be
extracted from the three-point correlation function%
\begin{eqnarray}
&&\Pi _{\mu \nu }(p,p^{\prime })=i^{2}\int d^{4}xd^{4}ye^{ip^{\prime
}y}e^{-ipx}\langle 0|\mathcal{T}\{J_{\mu }^{J/\psi }(y)  \notag \\
&&\times J_{\nu }^{J/\psi }(0)J^{\dagger }(x)\}|0\rangle ,  \label{eq:CF2}
\end{eqnarray}%
where $J_{\mu }^{J/\psi }(x)\ $is the interpolating currents for the $J/\psi
$ meson. The $J(x)$ is given by Eq.\ (\ref{eq:CR1}), while for $J_{\mu
}^{J/\psi }(x)$ we use%
\begin{equation}
J_{\mu }^{J/\psi }(x)=\overline{c}_{i}(x)\gamma _{\mu }c_{i}(x),
\label{eq:CR2}
\end{equation}%
where $i=1,2,3$ are the color indices. The $4$-momentum of the tetraquark $%
X_{\mathrm{4c}}$ is $p$, whereas momenta of the $J/\psi $ mesons are $%
p^{\prime }$ and $q=p-p^{\prime }$, respectively.

We follow the standard prescriptions of the sum rule method and express the
correlation function $\Pi _{\mu \nu }(p,p^{\prime })$ in terms of involved
particles' phenomenological parameters. Isolating the ground-state
contribution to the correlation function (\ref{eq:CF2}) from effects of
higher resonances and continuum states,\ for the physical side of the sum
rule $\Pi _{\mu \nu }^{\mathrm{Phys}}(p,p^{\prime })$, we get%
\begin{eqnarray}
&&\Pi _{\mu \nu }^{\mathrm{Phys}}(p,p^{\prime })=\frac{\langle 0|J_{\mu
}^{J/\psi }|J/\psi (p^{\prime })\rangle }{p^{\prime 2}-m_{1}^{2}}\frac{%
\langle 0|J_{\nu }^{J/\psi }|J/\psi (q)\rangle }{q^{2}-m_{1}^{2}}  \notag \\
&&\times \langle J/\psi (p^{\prime })J/\psi (q)|X_{\mathrm{4c}}(p)\rangle
\frac{\langle X_{\mathrm{4c}}(p)|J^{\dagger }|0\rangle }{p^{2}-m^{2}}+\cdots
,  \label{eq:CF3}
\end{eqnarray}%
with $m_{1}$ being the mass of the $J/\psi $ meson.

The function $\Pi _{\mu \nu }^{\mathrm{Phys}}(p,p^{\prime })$ can be
simplified by employing the matrix elements of the tetraquark $X_{\mathrm{4c}%
}$ and $J/\psi $ meson. The matrix element of $X_{\mathrm{4c}}$ is given by
Eq.\ (\ref{eq:ME1}), whereas for $\langle 0|J_{\mu }^{J/\psi }|J/\psi
(p)\rangle $ we use
\begin{equation}
\langle 0|J_{\mu }^{J/\psi }|J/\psi (p)\rangle =f_{1}m_{1}\varepsilon _{\mu
}(p),  \label{eq:ME2}
\end{equation}%
where $f_{1}$ and $\varepsilon _{\mu }$ are the decay constant and
polarization vector of the $J/\psi $ meson, respectively. We also model the
vertex $\langle J/\psi (p^{\prime })J/\psi (q)|X_{\mathrm{4c}}(p)\rangle $
by the expression%
\begin{eqnarray}
&&\langle J/\psi (p^{\prime })J/\psi (q)|X_{\mathrm{4c}}(p)\rangle
=g_{1}(q^{2})\left[ q\cdot p^{\prime }\varepsilon ^{\ast }(p^{\prime })\cdot
\varepsilon ^{\ast }(q)\right.  \notag \\
&&\left. -q\cdot \varepsilon ^{\ast }(p^{\prime })p^{\prime }\cdot
\varepsilon ^{\ast }(q)\right] ,  \label{eq:ME3}
\end{eqnarray}%
which has the gauge-invariant form.

After these transformations $\Pi _{\mu \nu }^{\mathrm{Phys}}(p,p^{\prime })$
is given by the formula%
\begin{eqnarray}
&&\Pi _{\mu \nu }^{\mathrm{Phys}}(p,p^{\prime })=g_{1}(q^{2})\frac{%
fmf_{1}^{2}m_{1}^{2}}{\left( p^{2}-m^{2}\right) \left( p^{\prime
2}-m_{1}^{2}\right) (q^{2}-m_{1}^{2})}  \notag \\
&&\times \left[ \frac{1}{2}\left( m^{2}-m_{1}^{2}-q^{2}\right) g_{\mu \nu
}-q_{\mu }p_{\nu }^{\prime }\right] +\cdots ,  \label{eq:CorrF5}
\end{eqnarray}%
where the ellipses stand for contributions of higher resonances and
continuum states. The correlator Eq.\ (\ref{eq:CorrF5}) contains different
Lorentz structures, which may be used to construct the sum rule for $%
g_{1}(q^{2})$. We choose to work with the term $\sim g_{\mu \nu }$ and
denote the relevant invariant amplitude by $\Pi ^{\mathrm{Phys}%
}(p^{2},p^{\prime 2},q^{2})$.

The correlation function $\Pi _{\mu \nu }(p,p^{\prime })$ calculated in
terms of heavy quark propagators reads
\begin{eqnarray}
&&\Pi _{\mu \nu }^{\mathrm{OPE}}(p,p^{\prime })=-2i^{2}\int
d^{4}xd^{4}ye^{ip^{\prime }y}e^{-ipx}  \notag \\
&&\times \left\{ \mathrm{Tr}\left[ \gamma _{\mu }S_{c}^{ib}(y-x)\gamma
_{\alpha }\widetilde{S}_{c}^{ja}(-x){}\gamma _{\nu }\widetilde{S}%
_{c}^{bj}(x)\gamma ^{\alpha }S_{c}^{ai}(x-y)\right] \right.  \notag \\
&&\left. -\mathrm{Tr}\left[ \gamma _{\mu }S_{c}^{ia}(y-x)\gamma _{\alpha }%
\widetilde{S}_{c}^{jb}(-x){}\gamma _{\nu }\widetilde{S}_{c}^{bj}(x)\gamma
^{\alpha }S_{c}^{ai}(x-y)\right] \right\}.  \notag \\
&&  \label{eq:QCDside}
\end{eqnarray}

\begin{figure}[h]
\includegraphics[width=8.5cm]{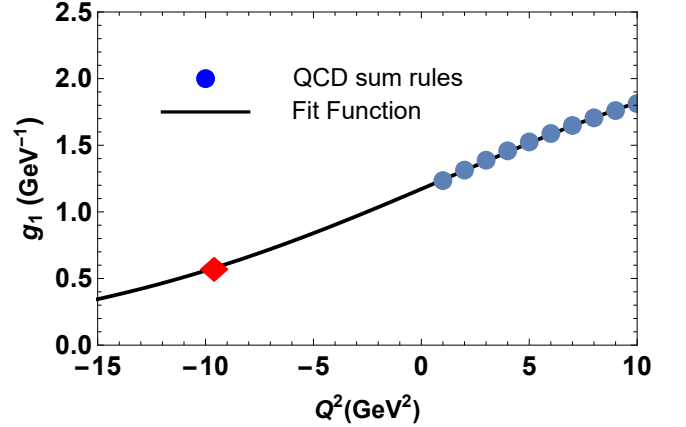}
\caption{The sum rule predictions and fit function for the strong coupling $%
g_{1}(Q^{2})$. The red diamond denotes the point $Q^{2}=-m_{1}^{2}$. }
\label{fig:Fit}
\end{figure}

\begin{table}[tbp]
\begin{tabular}{|c|c|}
\hline\hline
Parameters & Values (in $\mathrm{MeV}$ units) \\ \hline\hline
$m_1[m_{J/\psi}]$ & $3096.900 \pm 0.006$ \\
$f_1[f_{J/\psi}]$ & $409 \pm 15$ \\
$m_2[m_{\eta_c}]$ & $2983.9 \pm 0.4$ \\
$f_2[f_{\eta_c}]$ & $320 \pm 40$ \\
$m_3[m_{\chi _{c1}}]$ & $3510.67 \pm 0.05$ \\
$f_3[f_{\chi _{c1}}]$ & $344 \pm 27$ \\ \hline\hline
\end{tabular}%
\caption{Masses and decay constants of $\overline{c}c$ mesons, which have
been used in numerical computations. }
\label{tab:Param}
\end{table}
The invariant amplitude $\Pi ^{\mathrm{OPE}}(p^{2},p^{\prime 2},q^{2})$
which corresponds to the term $\sim g_{\mu \nu }$ in Eq.\ (\ref{eq:QCDside})
constitutes the QCD side of the sum rule. Having equated these two invariant
amplitudes, carried out the doubly Borel transformations over variables $%
p^{2}$ and $p^{\prime 2}$ and performed continuum subtraction, one finds the
sum rule for $g_{1}(q^{2})$
\begin{eqnarray}
&&g_{1}(q^{2})=\frac{2}{fmf_{1}^{2}m_{1}^{2}}\frac{q^{2}-m_{1}^{2}}{%
m^{2}-m_{1}^{2}-q^{2}}  \notag \\
&&\times e^{m^{2}/M_{1}^{2}}e^{m_{1}^{2}/M_{2}^{2}}\Pi (\mathbf{M}^{2},%
\mathbf{s}_{0},q^{2}).  \label{eq:SRCoup}
\end{eqnarray}%
Here, $\Pi (\mathbf{M}^{2},\mathbf{s}_{0},q^{2})$ is the amplitude $\Pi ^{%
\mathrm{OPE}}(p^{2},p^{\prime 2},q^{2})$ after the Borel transformation and
subtraction procedures: It can be expressed in terms of the spectral density
$\rho (s,s^{\prime },q^{2})$ calculated as an imaginary part of relevant
component of the correlation function $\Pi _{\mu \nu }^{\mathrm{OPE}%
}(p,p^{\prime })$,
\begin{eqnarray}
&&\Pi (\mathbf{M}^{2},\mathbf{s}_{0},q^{2})=\int_{16m_{c}^{2}}^{s_{0}}ds%
\int_{4m_{c}^{2}}^{s_{0}^{\prime }}ds^{\prime }\rho (s,s^{\prime },q^{2})
\notag \\
&&\times e^{-s/M_{1}^{2}}e^{-s^{\prime }/M_{2}^{2}},  \label{eq:SCoupl}
\end{eqnarray}%
where $\mathbf{M}^{2}=(M_{1}^{2},M_{2}^{2})$ and $\mathbf{s}%
_{0}=(s_{0},s_{0}^{\prime })$ are the Borel and continuum threshold
parameters, respectively.

The form factor $g_{1}(q^{2})$ depends on the masses and current couplings
(decay constant) of the tetraquark $X_{\mathrm{4c}}$ and the meson $J/\psi $
which appear in the numerical computations as input parameters. Their values
are moved to Table\ \ref{tab:Param}, which contains also spectroscopic
parameters of $\eta _{c}$ and $\chi _{c1}(1P)$ mesons required to
investigate two other decays of $X_{\mathrm{4c}}$. The masses all of the
mesons are borrowed from Ref.\ \cite{PDG:2022}. For the decay constant of
the meson $J/\psi $, we employ the experimental value reported in Ref.\ \cite%
{Kiselev:2001xa}. As $f_{\eta _{c}}$ and $f_{\chi _{c1}}$, we use
predictions made in Refs.\ \cite{Colangelo:1992cx,VeliVeliev:2012cc} on the
basis of the sum rule method, respectively.

To carry out numerical computations it is necessary also to choose the
working regions for the parameters $\mathbf{M}^{2}$ and $\mathbf{s}_{0}$.
The constraints imposed on $\mathbf{M}^{2}$ and $\mathbf{s}_{0}$ are
standard restrictions of the sum rule calculations and were explained in the
previous section. For $M_{1}^{2}$ and $s_{0}$, associated with the $X_{%
\mathrm{4c}}$ channel, we use the working windows from Eq.\ (\ref{eq:Wind1}%
). The parameters $(M_{2}^{2},\ s_{0}^{\prime })$ for the $J/\psi $ channel
are changed inside borders%
\begin{equation}
M_{2}^{2}\in \lbrack 4,5]~\mathrm{GeV}^{2},\ s_{0}^{\prime }\in \lbrack
12,13]~\mathrm{GeV}^{2}.  \label{eq:Wind3}
\end{equation}

It is known that the sum rule method leads to reliable predictions in the
deep-Euclidean region $q^{2}<0$. For our purposes, it is convenient to
introduce a new variable $Q^{2}=-q^{2}$ and denote the obtained function by $%
g_{1}(Q^{2})$. A range of $Q^{2}$ studied by the sum rule analysis covers
the region $Q^{2}=1-10~\mathrm{GeV}^{2}$. The results of calculations are
plotted in Fig.\ \ref{fig:Fit}. But the width of the decay $X_{\mathrm{4c}%
}\rightarrow J/\psi J/\psi $ is determined by the form factor $g_{1}(q^{2})$
at the mass shell $q^{2}=m_{1}^{2}$. Stated differently, one has to find $%
g_{1}(Q^{2}=-m_{1}^{2})$.

To solve this problem, we use a fit function $\mathcal{G}_{1}(Q^{2})$ that
at momenta $Q^{2}>0$ gives the same values as the sum rule calculations, but
can be extrapolated to the region of $Q^{2}<0$. In this paper, we employ the
functions $\mathcal{G}_{i}(Q^{2}),\ i=1,2,3$
\begin{equation}
\mathcal{G}_{i}(Q^{2})=\mathcal{G}_{i}^{0}\mathrm{\exp }\left[ c_{i}^{1}%
\frac{Q^{2}}{m^{2}}+c_{i}^{2}\left( \frac{Q^{2}}{m^{2}}\right) ^{2}\right] ,
\label{eq:FitF}
\end{equation}%
with parameters $\mathcal{G}_{i}^{0}$, $c_{i}^{1}$ and $c_{i}^{2}$.

Calculations prove that $\mathcal{G}_{1}^{0}=1.17~\mathrm{GeV}^{-1}$, $%
c_{1}^{1}=2.55$, and $c_{1}^{2}=-2.79$ give nice agreement with the sum
rule's data for $g_{1}(Q^{2})$ shown in Fig.\ \ref{fig:Fit}. At the mass
shell $q^{2}=m_{1}^{2}$ the function $\mathcal{G}_{1}(Q^{2})$ is equal to
\begin{equation}
g_{1}\equiv \mathcal{G}_{1}(-m_{1}^{2})=(5.8\pm 1.2)\times 10^{-1}\ \mathrm{%
GeV}^{-1}.
\end{equation}%
The partial width of the process $X_{\mathrm{4c}}\rightarrow J/\psi J/\psi $
can be obtained by employing the following expression
\begin{equation}
\Gamma \left[ X_{\mathrm{4c}}\rightarrow J/\psi J/\psi \right] =g_{1}^{2}%
\frac{\lambda }{8\pi }\left( \frac{m_{1}^{4}}{m^{2}}+\frac{2\lambda ^{2}}{3}%
\right) ,  \label{eq:PartDW}
\end{equation}%
where $\lambda =\lambda (m,m_{1},m_{1})$ and
\begin{equation}
\lambda (a,b,c)=\frac{\sqrt{%
a^{4}+b^{4}+c^{4}-2(a^{2}b^{2}+a^{2}c^{2}+b^{2}c^{2})}}{2a}.
\end{equation}

Then it is easy to find
\begin{equation}
\Gamma \left[ X_{\mathrm{4c}}\rightarrow J/\psi J/\psi \right] =(43\pm 13)~%
\mathrm{MeV}.  \label{eq:DW1}
\end{equation}

\begin{table}[tbp]
\begin{tabular}{|c|c|c|c|}
\hline\hline
$i$ & Channels & $g_{i}~(\mathrm{GeV}^{-1})$ & $\Gamma_{i}~(\mathrm{MeV})$
\\ \hline
$1$ & $X_{\mathrm{4c}}\to J/\psi J/\psi$ & $(5.8 \pm 1.2)\times 10^{-1}$ & $%
43 \pm 13$ \\
$2$ & $X_{\mathrm{4c}}\to \eta_{c}\eta_{c}$ & $(2.9 \pm 0.6)\times 10^{-1} $
& $51 \pm 15$ \\
$3$ & $X_{\mathrm{4c}} \to \eta_{c}\chi_{c1}(1P)$ & $10.9 \pm 2.8^{\star}$ &
$16 \pm 6 $ \\ \hline\hline
\end{tabular}%
\caption{Decay channels of the tetraquark $X_{\mathrm{4c}}$, strong
couplings $g_{i}$, and partial widths $\Gamma_{i}$. The coupling $g_3$
marked by a star is dimensionless.}
\label{tab:Channels}
\end{table}


\section{Processes $X_{\mathrm{4c}}\rightarrow \protect\eta _{c}\protect\eta %
_{c}$ and $X_{\mathrm{4c}}\rightarrow \protect\eta _{c}\protect\chi %
_{c1}(1P) $}

\label{sec:Decays2}

The decays $X_{\mathrm{4c}}\rightarrow \eta _{c}\eta _{c}$ and $X_{\mathrm{4c%
}}\rightarrow \eta _{c}\chi _{c1}(1P)$ can be explored in a similar manner.
The strong coupling $g_{2}$ that describes the vertex $X_{\mathrm{4c}}\eta
_{c}\eta _{c}$ can be extracted from the correlation function
\begin{eqnarray}
&&\Pi (p,p^{\prime })=i^{2}\int d^{4}xd^{4}ye^{ip^{\prime }y}e^{-ipx}\langle
0|\mathcal{T}\{J^{\eta _{c}}(y)  \notag \\
&&\times J^{\eta _{c}}(0)J^{\dagger }(x)\}|0\rangle,  \label{eq:CF4}
\end{eqnarray}%
where the current $J^{\eta _{c}}(x)$ is
\begin{equation}
J^{\eta _{c}}(x)=\overline{c}_{i}(x)i\gamma _{5}c_{i}(x).  \label{eq:CR3}
\end{equation}

Separating from each other the ground-state contribution and effects of
higher resonances and continuum states, we write the correlation function (%
\ref{eq:CF4}) in the following form%
\begin{eqnarray}
&&\Pi ^{\mathrm{Phys}}(p,p^{\prime })=\frac{\langle 0|J^{\eta _{c}}|\eta
_{c}(p^{\prime })\rangle }{p^{\prime 2}-m_{2}^{2}}\frac{\langle 0|J^{\eta
_{c}}|\eta _{c}(q)\rangle }{q^{2}-m_{2}^{2}}  \notag \\
&&\times \langle \eta _{c}(p^{\prime })\eta _{c}(q)|X_{\mathrm{4c}%
}(p)\rangle \frac{\langle X_{\mathrm{4c}}(p)|J^{\dagger }|0\rangle }{%
p^{2}-m^{2}}+\cdots,  \label{eq:CF5}
\end{eqnarray}%
where $m_{2}$ is the mass of the $\eta _{c}$ meson. We define the vertex
composed of a scalar and two pseudoscalar particles by means of the formula
\begin{equation}
\langle \eta _{c}(p^{\prime })\eta _{c}(q)|X_{\mathrm{4c}}(p)\rangle
=g_{2}(q^{2})p\cdot p^{\prime }.  \label{eq:ME5}
\end{equation}%
To express the correlator $\Pi ^{\mathrm{Phys}}(p,p^{\prime })$ in terms of
physical parameters of particles $X_{\mathrm{4c}}$ and $\eta _{c}$, we use
the matrix element Eq.\ (\ref{eq:ME1}) and
\begin{equation}
\langle 0|J^{\eta _{c}}|\eta _{c}\rangle =\frac{f_{2}m_{2}^{2}}{2m_{c}},
\label{eq:ME4}
\end{equation}%
with $f_{2}$ being the decay constant of the $\eta _{c}$ meson. Then, the
correlation function $\Pi ^{\mathrm{Phys}}(p,p^{\prime })$ takes the form%
\begin{eqnarray}
&&\Pi ^{\mathrm{Phys}}(p,p^{\prime })=g_{2}(q^{2})\frac{fmf_{2}^{2}m_{2}^{4}%
}{4m_{c}^{2}\left( p^{2}-m^{2}\right) \left( p^{\prime 2}-m_{2}^{2}\right) }
\notag \\
&&\times \frac{m^{2}+m_{2}^{2}-q^{2}}{2(q^{2}-m_{2}^{2})}+\cdots .
\label{eq:CF6}
\end{eqnarray}%
The function $\Pi ^{\mathrm{Phys}}(p,p^{\prime })$ has a Lorentz structure
that is proportional to $\mathrm{I}$, hence rhs of Eq.\ (\ref{eq:CF6}) is
the corresponding invariant amplitude $\widetilde{\Pi }^{\mathrm{Phys}%
}(p^{2},p^{\prime 2},q^{2})$.

Using the heavy quark propagators, we can find the QCD side of the sum rule
\begin{eqnarray}
&&\Pi ^{\mathrm{OPE}}(p,p^{\prime })=2i^{2}\int d^{4}xd^{4}ye^{ip^{\prime
}y}e^{-ipx}  \notag \\
&&\times \left\{ \mathrm{Tr}\left[ \gamma _{5}S_{c}^{ia}(y-x)\gamma _{\alpha
}\widetilde{S}_{c}^{jb}(-x){}\gamma _{5}\widetilde{S}_{c}^{bj}(x)\gamma
^{\alpha }S_{c}^{ai}(x-y)\right] \right.  \notag \\
&&\left. -\mathrm{Tr}\left[ \gamma _{5}S_{c}^{ia}(y-x)\gamma _{\alpha }%
\widetilde{S}_{c}^{jb}(-x){}\gamma _{5}\widetilde{S}_{c}^{aj}(x)\gamma
^{\alpha }S_{c}^{bi}(x-y)\right] \right\}.  \notag \\
&&  \label{eq:QCDside2}
\end{eqnarray}%
The sum rule for the strong form factor $g_{2}(q^{2})$ equals to%
\begin{eqnarray}
&&g_{2}(q^{2})=\frac{8m_{c}^{2}}{fmf_{2}^{2}m_{2}^{4}}\frac{q^{2}-m_{2}^{2}}{%
m^{2}+m_{2}^{2}-q^{2}}  \notag \\
&&\times e^{m^{2}/M_{1}^{2}}e^{m_{2}^{2}/M_{2}^{2}}\widetilde{\Pi }(\mathbf{M%
}^{2},\mathbf{s}_{0},q^{2}),  \label{eq:SRCoup2}
\end{eqnarray}%
with $\widetilde{\Pi }(\mathbf{M}^{2},\mathbf{s}_{0},q^{2})$ being the
invariant amplitude $\widetilde{\Pi }^{\mathrm{OPE}}(p^{2},p^{\prime
2},q^{2})$ corresponding to the correlator $\Pi ^{\mathrm{OPE}}(p,p^{\prime
})$ after the Borel transformations and continuum subtractions.

We carry out numerical computations using Eq.\ (\ref{eq:SRCoup2}),
parameters of the meson $\eta _{c}$ from Table\ \ref{tab:Param}, and working
regions for $\mathbf{M}^{2}$ and $\mathbf{s}_{0}$. The Borel and continuum
subtraction parameters $M_{1}^{2}$ and $s_{0}$ in the $X_{\mathrm{4c}}$
channel is chosen as in Eq.\ (\ref{eq:Wind1}), whereas for $M_{2}^{2}$ and $%
s_{0}^{\prime }$ which correspond to the $\eta _{c}$ channel, we employ
\begin{equation}
M_{2}^{2}\in \lbrack 3.5,4.5]~\mathrm{GeV}^{2},\ s_{0}^{\prime }\in \lbrack
11,12]~\mathrm{GeV}^{2}.  \label{eq:Wind4}
\end{equation}

The interpolating function $\mathcal{G}_{2}(Q^{2})$ has the parameters $%
\mathcal{G}_{2}^{0}=0.65~\mathrm{GeV}^{-1}$, $c_{2}^{1}=3.19$, and $%
c_{2}^{2}=-3.34$. For the strong coupling $g_{2}$, we get
\begin{equation}
g_{2}\equiv \mathcal{G}_{2}(-m_{2}^{2})=(2.9\pm 0.6)\times 10^{-1}\ \mathrm{%
GeV}^{-1}.
\end{equation}%
The width of the process $X_{\mathrm{4c}}\rightarrow \eta _{c}\eta _{c}$ is
determined by means of the formula%
\begin{equation}
\Gamma \left[ X_{\mathrm{4c}}\rightarrow \eta _{c}\eta _{c}\right] =g_{2}^{2}%
\frac{m_{2}^{2}\widetilde{\lambda }}{8\pi }\left( 1+\frac{\widetilde{\lambda
}^{2}}{m_{2}^{2}}\right),  \label{eq:PDw2}
\end{equation}%
where $\widetilde{\lambda }=\lambda (m,m_{2},m_{2})$. Finally, we obtain
\begin{equation}
\Gamma \left[ X_{\mathrm{4c}}\rightarrow \eta _{c}\eta _{c}\right] =(51\pm
15)~\mathrm{MeV}.  \label{eq:DW2}
\end{equation}

Treatment of the $P$-wave decay $X_{\mathrm{4c}}\rightarrow \eta _{c}\chi
_{c1}(P)$ does not generate additional technical details, and is performed
in a usual manner. The three-point correlator to be considered in this case
is
\begin{eqnarray}
&&\Pi _{\mu }(p,p^{\prime })=i^{2}\int d^{4}xd^{4}ye^{ip^{\prime
}y}e^{-ipx}\langle 0|\mathcal{T}\{J_{\mu }^{\chi _{c1}}(y)  \notag \\
&&\times J^{\eta _{c}}(0)J^{\dagger }(x)\}|0\rangle,  \label{eq:CF7}
\end{eqnarray}%
where $J_{\mu }^{\chi _{c1}}(y)$ is the interpolating current for the meson $%
\chi _{c1}(1P)$%
\begin{equation}
J_{\mu }^{\chi _{c1}}(y)=\overline{c}_{j}(x)\gamma _{5}\gamma _{\mu
}c_{j}(x).  \label{eq:Curr1}
\end{equation}

In terms of the physical parameters of the particles the correlation
function has the form%
\begin{eqnarray}
&&\Pi _{\mu }^{\mathrm{Phys}}(p,p^{\prime })=g_{3}(q^{2})\frac{%
fmf_{2}m_{2}^{2}f_{3}m_{3}}{2m_{c}\left( p^{2}-m^{2}\right) \left( p^{\prime
2}-m_{3}^{2}\right) }  \notag \\
&&\times \frac{1}{q^{2}-m_{2}^{2}}\left[ \frac{m^{2}-m_{3}^{2}-q^{2}}{%
2m_{3}^{2}}p_{\mu }^{\prime }-q_{\mu }\right] +\cdots .  \label{eq:CF8}
\end{eqnarray}%
In Eq.\ (\ref{eq:CF8}) $m_{3}$ and $f_{3}$ are the mass and decay constant
of the meson $\chi _{c1}(1P)$. To derive the correlator $\Pi _{\mu }^{%
\mathrm{Phys}}(p,p^{\prime })$, we have used the known matrix elements of
the tetraquark $X_{\mathrm{4c}}$ and meson $\eta _{c}$, as well as new
matrix elements
\begin{equation}
\langle 0|J_{\mu }^{\chi _{c1}}|\chi _{c1}(p^{\prime })\rangle
=f_{3}m_{3}\varepsilon _{\mu }^{\ast }(p^{\prime }),
\end{equation}%
and
\begin{equation}
\langle \eta _{c}(q)\chi _{c1}(p^{\prime })|X_{\mathrm{4c}}(p)\rangle
=g_{3}(q^{2})p\cdot \varepsilon ^{\ast }(p^{\prime }),
\end{equation}%
where $\varepsilon _{\mu }^{\ast }(p^{\prime })$ is the polarization vector
of $\chi _{c1}(1P)$.

The QCD side $\Pi _{\mu }^{\mathrm{OPE}}(p,p^{\prime })$ is given by the
formula
\begin{eqnarray}
&&\Pi _{\mu }^{\mathrm{OPE}}(p,p^{\prime })=2i^{3}\int
d^{4}xd^{4}ye^{ip^{\prime }y}e^{-ipx}  \notag \\
&&\times \left\{ \mathrm{Tr}\left[ \gamma _{\mu }\gamma
_{5}S_{c}^{ia}(y-x)\gamma _{\alpha }\widetilde{S}_{c}^{jb}(-x){}\gamma _{5}%
\widetilde{S}_{c}^{bj}(x)\gamma ^{\alpha }S_{c}^{ai}(x-y)\right] \right.
\notag \\
&&\left. -\mathrm{Tr}\left[ \gamma _{\mu }\gamma _{5}S_{c}^{ia}(y-x)\gamma
_{\alpha }\widetilde{S}_{c}^{jb}(-x){}\gamma _{5}\widetilde{S}%
_{c}^{aj}(x)\gamma ^{\alpha }S_{c}^{bi}(x-y)\right] \right\}.  \notag \\
&&  \label{eq:CF9}
\end{eqnarray}

The sum rule for $g_{3}(q^{2})$ is derived using the invariant amplitudes
corresponding to terms $\sim p_{\mu }^{\prime }$ in both $\Pi _{\mu }^{%
\mathrm{Phys}}(p,p^{\prime })$ and $\Pi _{\mu }^{\mathrm{OPE}}(p,p^{\prime
}) $. In numerical analysis, $M_{2}^{2}$ and $s_{0}^{\prime }$ in the $\chi
_{c1}$ channel are chosen in the following way
\begin{equation}
M_{2}^{2}\in \lbrack 4,5]~\mathrm{GeV}^{2},\ s_{0}^{\prime }\in \lbrack
13,14]~\mathrm{GeV}^{2}.  \label{eq:Wind5}
\end{equation}%
For the parameters of the fit function $\mathcal{G}_{3}(Q^{2})$, we get $%
\mathcal{G}_{3}^{0}=24.08$, $c_{3}^{1}=2.98$, and $c_{3}^{2}=-4.26$. Then,
the strong coupling $g_{3}$ is equal to
\begin{equation}
g_{3}\equiv \mathcal{G}_{3}(-m_{2}^{2})=10.9\pm 2.8.
\end{equation}

The width of the decay $X_{\mathrm{4c}}\rightarrow \eta _{c}\chi _{c1}(P)$ \
can be calculated by means of the expression
\begin{equation}
\Gamma \left[ X_{\mathrm{4c}}\rightarrow \eta _{c}\chi _{c1}(P)\right]
=g_{3}^{2}\frac{\widehat{\lambda }^{3}}{24\pi m_{3}^{2}},  \label{eq:DW3}
\end{equation}%
where $\widehat{\lambda }=\lambda (m,m_{2},m_{3})$. For the width of this
process, we obtain the estimate:
\begin{equation}
\Gamma \left[ X_{\mathrm{4c}}\rightarrow \eta _{c}\chi _{c1}(P)\right]
=(16\pm 6)~\mathrm{MeV}.  \label{eq:DW4}
\end{equation}%
The widths all of three decays are collected in Table\ \ref{tab:Channels}.
Based on these results, it is not difficult to find that
\begin{equation}
\Gamma _{\mathrm{4c}}=(110\pm 21)~\mathrm{MeV},  \label{eq:FW}
\end{equation}%
which nicely agrees with CMS datum $\Gamma _{1}^{\mathrm{CMS}}$.


\section{Discussion and concluding notes}

\label{sec:Disc}

In the present article, we have performed detailed analysis of the
tetraquark $X_{\mathrm{4c}}$ by calculating the mass $m$ and full width $%
\Gamma _{\mathrm{4c}}$ of this scalar diquark-antidiquark state. Our
findings are in agreements with the experimental data $m_{1}^{\mathrm{CMS}%
}=(6552\pm 10\pm 12)~\mathrm{MeV}$ and $\Gamma _{1}^{\mathrm{CMS}%
}=(124_{-26}^{+32}\pm 33)~\mathrm{MeV}$ of the CMS Collaboration. The mass
of $X_{\mathrm{4c}}$ is compatible also with $m_{1}^{\mathrm{ATL}}$ if one
takes into account existing experimental and theoretical errors. We have
interpreted the ground-level $1S$ tetraquark $X_{\mathrm{4c}}$ built of
axial-vector constituents as the resonance $X(6600)$.

The partial width of the decay $X_{\mathrm{4c}}\rightarrow \eta _{c}\eta
_{c} $ is comparable with $\Gamma \left[ X_{\mathrm{4c}}\rightarrow J/\psi
J/\psi \right] $. The new fully charmed resonances were observed in the di-$%
J/\psi $ mass distribution through $4\mu $ final states. It is known that
decays to lepton pairs $e^{+}e^{-}$ and $\mu ^{+}\mu ^{-}$ are among
important modes of the $J/\psi $ meson \cite{PDG:2022}. But, the $\eta _{c}$
meson's main channels are decays to hadronic resonances, for example, to $%
\rho \rho $ mesons. Naturally, the process $X_{\mathrm{4c}}\rightarrow \eta
_{c}\eta _{c} $ could not be seen in $4\mu $ events.

There are numerous publications, in which properties of the tetraquark $X_{%
\mathrm{4c}}$ were studied using various methods (for complete list of
relevant publications see, Ref.\ \cite{Faustov:2022mvs}). These
investigations intensified after discovery of resonances $X(6200)$, $X(6600)$%
, $X(6900)$ and $X(7300)$. Comparing our result for the mass of $X_{\mathrm{%
4c}}$ with $(6.46\pm 0.16)~\mathrm{GeV}$ and $6.46_{-0.17}^{+0.13}~\mathrm{%
GeV}$ from Refs.\ \cite{Chen:2016jxd,Zhang:2020xtb}, we see that though $m$
exceeds them, within ambiguities of calculations all predictions are
comparable with each other. But what is more important, decays to $J/\psi
J/\psi $ pairs are kinematically allowed channels for these structures.

The first resonance $X(6200)$ in the list of the fully charmed states may be
a manifestation of the hadronic molecule $\eta _{c}\eta _{c}$ in the $J/\psi
J/\psi $ spectrum. But to be detected the mass of $\eta _{c}\eta _{c}$ must
exceed the di-$J/\psi $ threshold $\simeq 6195~\mathrm{MeV}$. In Ref.\ \cite%
{Albuquerque:2020hio} the authors predicted $M_{\eta _{c}\eta _{c}}=6029\pm
198~\mathrm{MeV}$ that in upper limit overshoots the di-$J/\psi $ threshold.
Alternatively, appearance of the near-threshold state $X(6200)$ may be
explained by coupled-channel effects \cite{Dong:2020nwy}.

The next structure, $X(6900)$, can be considered in the diquark-antidiquark
model provided it composed of pseudoscalar components. In fact, the mass of
such tetraquark was estimated around $(6.82\pm 0.18)~\mathrm{GeV}$ and $%
(6.80\pm 0.27)~\mathrm{GeV}$ in Refs.\ \cite%
{Chen:2016jxd,Albuquerque:2020hio}, respectively. The hadronic molecule $%
\chi _{c0}\chi _{c0}$ with the mass $\simeq 6.93~\mathrm{GeV}$ is an
alternative candidate to the resonance $X(6900)$ \cite{Albuquerque:2020hio}.

More detailed analyses of assumptions about a diquark-antidiquark or
hadronic molecule nature of the resonances $X(6200)$ and $X(6900)$ were
performed in our articles \cite{Agaev:2023gaq,Agaev:2023ruu}. In these
works, we applied the sum rule method to investigate the diquark-antidiquark
state $T_{\mathrm{4c}}$ built of pseudoscalar constitutes $c_{a}^{T}Cc_{b}$
and $\overline{c}_{a}C\overline{c}_{b}^{T}$, as well as hadronic molecules $%
\eta _{c}\eta _{c}$ and $\chi _{c0}\chi _{c0}$ . In Ref. \cite{Agaev:2023ruu}
it was demonstrated that the molecule $\eta _{c}\eta _{c}$ with the mass $%
(6264\pm 50)~\mathrm{MeV}$ and full width $(320\pm 72)~\mathrm{MeV}$ is a
natural candidate to the resonance $X(6200)$. Our prediction for the mass of
this molecule is larger than $M_{\eta _{c}\eta _{c}}$, but has some
overlapping region with it.

The mass of the tetraquark $T_{\mathrm{4c}}$ amounts to $(6928\pm 50)~%
\mathrm{MeV}$ and is compatible with previous sum rule predictions and
relevant LHCb-ATLAS-CMS data, especially with the result of the CMS
Collaboration for $X(6900)$ \cite{Agaev:2023gaq}. The full width of $T_{%
\mathrm{4c}}$ was evaluated by taking into account its allowed decay
channels and found equal to $(128\pm 22)~\mathrm{MeV}$ in agreement with the
CMS measurements. The parameters of the molecule $\chi _{c0}\chi _{c0}$ are
equal to $(6954\pm 50)~\mathrm{MeV}$ and $(138\pm 18)~\mathrm{MeV}$,
respectively \cite{Agaev:2023ruu}. It may also be interpreted as a resonance
$X(6900)$, or considered as its part in the tetraquark-molecule mixing model.

As is seen, though diquark-antidiquark states and hadronic molecules have
different internal organizations, both of them may be used to model $X$
resonances. Such "universality" of the $X$ structures is connected mainly
with errors of measurements reported by different collaborations. To make a
choice between different models for $X$ particles, one needs more precise
data on their parameters.

The heaviest state $X(7300)$ from this list is presumably a radially excited
$X_{\mathrm{4c}}(2S)$ tetraquark. An argument in favor of such assumption
came from the ATLAS Collaboration, which fixed the resonances $X(6600)$ and $%
X(7300)$ in the $J/\psi J/\psi $ and $J/\psi \psi ^{\prime }$ mass
distributions, respectively. In other words%
\begin{eqnarray}
X(7300) &\rightarrow &J/\psi \psi ^{\prime },  \notag \\
X(6600) &\rightarrow &J/\psi J/\psi ,
\end{eqnarray}%
are decay modes of these resonances. The mass gap between $\psi ^{\prime }$
and $J/\psi $ is around $590$ $\mathrm{MeV}$, whereas for $X(7300)$ and $%
X(6600)$ the mass difference equals to $600$ $\mathrm{MeV}$ (ATLAS) and $735$
$\mathrm{MeV}$ (CMS). Then, it is natural to suppose that $X(7300)$ is the
first radially excited state of $X(6600)$. Originally, similar hypothesis
was made in Ref.\ \cite{Maiani:2014}, while considering the main decay
channels of the resonances $Z_{c}(3900)$ and $Z_{c}(4330)$:
\begin{eqnarray}
Z_{c}(4330) &\rightarrow &\psi ^{\prime }\pi ,  \notag \\
Z_{c}(3900) &\rightarrow &J/\psi \pi .
\end{eqnarray}%
It was supposed that $Z_{c}(4330)$ is first radial excitation of the
tetraquark $Z_{c}(3900)$. This idea was later confirmed by calculations
carried out using the diquark-antidiquark model and sum rule method \cite%
{Wang:2014vha,Agaev:2017tzv}. In light of this analysis the assumption about
$2S$ excited nature of $X(7300)$ looks plausible. Results of our
investigations seem support this assumption and will be reported very soon.

We have calculated also the mass of the fully beauty scalar state $X_{%
\mathrm{4b}}$. It turned out that, its mass $m^{\prime }=(18540\pm 50)~%
\mathrm{MeV}$ is smaller than the $\eta _{b}\eta _{b}$ threshold, and hence $%
X_{\mathrm{4b}}$ does not decay to a pair of hidden-bottom mesons and cannot
be observed in $\eta _{b}\eta _{b}$ or $\Upsilon (1S)\Upsilon (1S)$ mass
distributions. The stability of $X_{\mathrm{4b}}$ in these channels was
already predicted in Refs.\ \cite{Berezhnoy:2011xn,Chen:2016jxd}. Its
transformation to ordinary mesons can proceed through subprocesses $%
\overline{b}b\rightarrow \overline{q}q(\overline{s}s)$ and $\overline{b}%
b\rightarrow 2g(3g)$ that result in the decay $X_{\mathrm{4b}}\rightarrow
B^{+}B^{-}$ and other similar processes \cite{Becchi:2020mjz}. The weak
leptonic and nonleptonic decays of $X_{\mathrm{4b}}$ are also among its
possible transitions to conventional mesons.

It is clear that controversial character of conclusions about nature of the
fully heavy resonances is connected with different models and schemes
employed for their investigations. In some of these articles, for instance, $%
X_{\mathrm{4b}}$ can decay to a pair of pseudoscalar mesons $\eta _{b}\eta
_{b}$, but is stable against $\Upsilon (1S)\Upsilon (1S)$ mode, whereas in
other publications $X_{\mathrm{4b}}$ is stable in both of these channels. In
the case of fully charmed states a same resonance due to large experimental
errors, may be interpreted within both the molecule and diquark-antidiquark
models.

We would like to emphasize that a large part of conclusions about the
ground-state and excited states $X_{\mathrm{4c}}$ and $X_{\mathrm{4b}}$ was
drawn using information on masses of these structures. In our view, in
scenarios with four-quark mesons one has to calculate also their widths,
otherwise statements made by relying only on the masses of these structures
remain not fully convincing.

\begin{widetext}

\appendix*

\section{ Heavy quark propagator $S_{Q}^{ab}(x)$ and spectral density $%
\protect\rho ^{\mathrm{pert.}}(s,\protect\alpha ,\protect\beta ,\protect%
\gamma )$}

\renewcommand{\theequation}{\Alph{section}.\arabic{equation}} \label{sec:App}

In the current article, for the heavy quark propagator $S_{Q}^{ab}(x)$ ($%
Q=c,\ b$), we employ
\begin{eqnarray}
&&S_{Q}^{ab}(x)=i\int \frac{d^{4}k}{(2\pi )^{4}}e^{-ikx}\Bigg \{\frac{\delta
_{ab}\left( {\slashed k}+m_{Q}\right) }{k^{2}-m_{Q}^{2}}-\frac{%
g_{s}G_{ab}^{\alpha \beta }}{4}\frac{\sigma _{\alpha \beta }\left( {\slashed %
k}+m_{Q}\right) +\left( {\slashed k}+m_{Q}\right) \sigma _{\alpha \beta }}{%
(k^{2}-m_{Q}^{2})^{2}}  \notag \\
&&+\frac{g_{s}^{2}G^{2}}{12}\delta _{ab}m_{Q}\frac{k^{2}+m_{Q}{\slashed k}}{%
(k^{2}-m_{Q}^{2})^{4}}+\cdots \Bigg \}.
\end{eqnarray}%
Here, we have used the notations
\begin{equation}
G_{ab}^{\alpha \beta }\equiv G_{A}^{\alpha \beta }\lambda _{ab}^{A}/2,\ \
G^{2}=G_{\alpha \beta }^{A}G_{A}^{\alpha \beta },\
\end{equation}%
where $G_{A}^{\alpha \beta }$ is the gluon field-strength tensor, and $%
\lambda ^{A}$ are the Gell-Mann matrices. The indices $A,B,C$ run in the
range $1,2,\ldots 8$.

The invariant amplitude $\Pi (M^{2},s_{0})$ obtained after the Borel
transformation and subtraction procedures is given by the expression%
\begin{equation*}
\Pi (M^{2},s_{0})=\int_{16m_{Q}^{2}}^{s_{0}}ds\rho ^{\mathrm{OPE}%
}(s)e^{-s/M^{2}},
\end{equation*}%
where the spectral density $\rho ^{\mathrm{OPE}}(s)$ is determined by the
formula
\begin{equation}
\rho ^{\mathrm{OPE}}(s)=\rho ^{\mathrm{pert.}}(s)+\langle \alpha
_{s}G^{2}/\pi \rangle \rho ^{\mathrm{Dim4}}(s).  \label{eq:A1}
\end{equation}%
The components $\rho ^{\mathrm{pert.}}(s)$ and $\rho ^{\mathrm{Dim4}}(s)$ of
the spectral density are
\begin{equation}
\rho ^{\mathrm{pert.(Dim4)}}(s)=\int_{0}^{1}d\alpha \int_{0}^{1-a}d\beta
\int_{0}^{1-a-\beta }d\gamma \rho ^{\mathrm{pert.(Dim4)}}(s,\alpha ,\beta
,\gamma ),\ \   \label{eq:A2}
\end{equation}%
where the variables $\alpha $, $\beta $, and $\gamma $ are Feynman
parameters.

The function $\rho ^{\mathrm{pert.}}(s,\alpha ,\beta ,\gamma )$ has the form
\begin{eqnarray}
&&\rho ^{\mathrm{pert.}}(s,\alpha ,\beta ,\gamma )=\frac{\Theta
(L_{1})N_{1}^{2}}{64\pi ^{6}N_{2}^{8}N_{3}^{5}(1-\gamma -\beta )^{2}}\left\{
-6m_{Q}^{4}(\beta +\gamma
-1)^{2}N_{2}^{4}N_{3}^{3}+m_{Q}^{2}N_{2}^{2}N_{3}\left\{ 3Ls\alpha \left[
N_{2}^{2}(N_{3}-L\alpha )\right. \right. \right.  \notag \\
&&+LN_{2}\alpha \gamma (-N_{3}(2N_{3}+(\beta +\gamma -1)^{2})+4N_{3}\alpha
(\beta +\gamma -1)-2\alpha ^{2}(L^{2}-2N_{3}))+L^{2}\alpha ^{2}\gamma
^{2}\left( N_{3}(N_{3}+(\gamma +\beta -1)^{2})\right.  \notag \\
&&\left. \left. -2N_{3}\alpha (\beta +\gamma -1)+\alpha
^{2}(L^{2}-2N_{3})\right) \right] +2N_{1}\left[ -LN_{3}^{2}\alpha -L\alpha
(\gamma (\beta +\gamma -1)+\alpha (\gamma +\beta -1)+\alpha ^{2})\right.
\notag \\
&&\times (\beta (\gamma +\beta -1)+\alpha (\beta +\gamma -1)+\alpha
^{2})+N_{3}\left( \gamma \beta (\beta +\gamma -1)^{2}+\alpha (\beta +\gamma
-1)^{2}+\alpha ^{2}(\beta +\gamma -1)(2\beta +2\gamma -1)\right.  \notag \\
&&\left. \left. \left. +4\alpha ^{3}(\beta +\gamma -1)+2\alpha ^{4}\right)
\right] \right\} -3L\alpha (L\alpha -N_{3})\left\{ 2L^{2}s^{2}\alpha
^{2}\gamma (N_{3}-L\alpha )(N_{2}-L\alpha \gamma )+2LN_{1}s\alpha \left[
2L^{2}\alpha ^{2}\gamma ^{2}(N_{3}-L\alpha )^{3}\right. \right.  \notag \\
&&+N_{2}^{2}(N_{3}+\gamma (\beta +\gamma -1)-\alpha L)+LN_{2}\alpha \gamma
(-3N_{3}-\gamma (\beta +\gamma -1)+3\alpha L)+N_{1}^{2}(-LN_{3}\alpha
+(\beta ^{2}+(\beta +\alpha )(\alpha +\gamma -1))  \notag \\
&&\left. \left. \left. \times (\gamma ^{2}+(\gamma +\alpha )(\alpha +\beta
-1))\right] \right\} \right\},
\end{eqnarray}

In expressions above, $\Theta (z)$ is the Unit Step function. We have used
also the following notations%
\begin{eqnarray}
&&N_{1}=s\alpha \beta \gamma \left[ \gamma ^{3}+2\gamma ^{2}(\beta +\alpha
-1)+\alpha (\beta +\alpha -1)+\gamma \left( 1+\beta ^{2}-3\alpha +2\alpha
^{2}\right. \right.  \notag \\
&&\left. \left. +\beta (-2+3\alpha )\right) \right] -m_{Q}^{2}\left[ \beta
\alpha ^{2}(\alpha +\beta -1)^{2}+\gamma ^{4}(\alpha +\beta )+\gamma \alpha
(\alpha +\beta -1)^{2}(2\beta +\alpha )\right.  \notag \\
&&\left. +2\gamma ^{3}(\beta ^{2}+\alpha (\alpha -1)+\beta (2\alpha
-1))+\gamma ^{2}(\beta ^{3}+\beta ^{2}(5\alpha -2)+\alpha (1-3\alpha
+2\alpha ^{2})+\beta (1-6\alpha +6\alpha ^{2}))\right] ,  \notag \\
&&N_{2}=\beta \alpha (\alpha +\beta -1)+\gamma ^{2}(\alpha +\beta )+\gamma
\left[ \beta ^{2}+\alpha (\alpha -1)+\beta (2\alpha -1)\right] ,  \notag \\
&&N_{3}=\gamma ^{2}+(\gamma +\alpha )(\beta +\alpha -1),\ \ L=\alpha +\beta
+\gamma -1,\ L_{1}=N_{1}/N_{2}^{2}.
\end{eqnarray}

\end{widetext}


\begin{thebibliography}{999}

\bibitem{Jaffe:1976ig} R.~L.~Jaffe,
Phys.\ Rev.\ D \textbf{15}, 267 (1977). 


\bibitem{Jaffe:1976yi} R.~L.~Jaffe, 
Phys.\ Rev.\ Lett.\ \textbf{38}, 195 (1977); \textbf{38}, 617(E) (1977).


\bibitem{Ader:1981db} J.~P.~Ader, J.~M.~Richard, and P.~Taxil,
Phys.\ Rev.\ D \textbf{25}, 2370 (1982).


\bibitem{Lipkin:1986dw} H.~J.~Lipkin,
Phys.\ Lett.\ B \textbf{172}, 242 (1986).


\bibitem{Zouzou:1986qh} S.~Zouzou, B.~Silvestre-Brac, C.~Gignoux, and
J.~M.~Richard, 
Z.\ Phys.\ C \textbf{30}, 457 (1986).


\bibitem{Carlson:1987hh} J.~Carlson, L.~Heller, and J.~A.~Tjon,
Phys.\ Rev.\ D \textbf{37}, 744 (1988).


\bibitem{Berezhnoy:2011xn} A.~V.~Berezhnoy, A.~V.~Luchinsky, and
A.~A.~Novoselov, 
Phys.\ Rev.\ D \textbf{86}, 034004 (2012).


\bibitem{Karliner:2016zzc} M.~Karliner, S.~Nussinov, and J.~L.~Rosner,
Phys.\ Rev.\ D \textbf{95}, 034011 (2017). 


\bibitem{Wu:2016vtq} J.~Wu, Y.~R.~Liu, K.~Chen, X.~Liu, and S.~L.~Zhu,
Phys.\ Rev.\ D \textbf{97}, 094015 (2018).


\bibitem{Chen:2016jxd} W.~Chen, H.~X.~Chen, X.~Liu, T.~G.~Steele, and
S.~L.~Zhu,
Phys.\ Lett.\ B \textbf{773}, 247 (2017).


\bibitem{Wang:2017jtz} Z.~G.~Wang,
Eur.\ Phys.\ J.\ C \textbf{77}, 432 (2017).


\bibitem{Richard:2017vry} J.~M.~Richard, A.~Valcarce, and J.~Vijande,
Phys.\ Rev.\ D \textbf{95}, 054019 (2017).


\bibitem{LHCb:2011kri} R.~Aaij \textit{et al.} (LHCb Collaboration),
Phys.\ Lett.\ B \textbf{707}, 52 (2012). 


\bibitem{CMS:2014cmt} V.~Khachatryan \textit{et al.} (CMS Collaboration),
JHEP \textbf{09}, 094 (2014). 


\bibitem{D0:2014vql} V.~M.~Abazov \textit{et al.} (D0 Collaboration),
Phys.\ Rev.\ D \textbf{90}, 111101 (2014). 


\bibitem{D0:2015dyx} V.~M.~Abazov \textit{et al.} (D0 Collaboration),
Phys.\ Rev.\ Lett. \textbf{116}, 082002 (2016).


\bibitem{CMS:2016liw} V.~Khachatryan \textit{et al.} (CMS Collaboration),
JHEP \textbf{05}, 013 (2017).


\bibitem{Aaij:2017ueg} R.~Aaij \textit{et al.} (LHCb Collaboration),
Phys.\ Rev.\ Lett.\ \textbf{119}, 112001 (2017).


\bibitem{Karliner:2017qjm} M.~Karliner and J.~L.~Rosner,
Phys.\ Rev.\ Lett.\ \textbf{119}, 202001 (2017).


\bibitem{Eichten:2017ffp} E.~J.~Eichten and C.~Quigg,
Phys.\ Rev.\ Lett.\ \textbf{119}, 202002 (2017).


\bibitem{Xing:2018bqt} Y.~Xing and R.~Zhu,
Phys.\ Rev.\ D \textbf{98}, 053005 (2018). 


\bibitem{Li:2018bkh} G.~Li, X.~F.~Wang, and Y.~Xing,
Eur.\ Phys.\ J.\ C \textbf{79}, 210 (2019). 


\bibitem{Agaev:2018khe} S.~S.~Agaev, K.~Azizi, B.~Barsbay, and H.~Sundu,
Phys.\ Rev.\ D \textbf{99}, 033002 (2019).


\bibitem{Agaev:2020mqq} S.~S.~Agaev, K.~Azizi, B.~Barsbay, and H.~Sundu,
Eur.\ Phys.\ J.\ A \textbf{57}, 106 (2021).


\bibitem{Agaev:2020dba} S.~S.~Agaev, K.~Azizi, B.~Barsbay, and H.~Sundu,
Eur.\ Phys.\ J.\ A \textbf{56}, 177 (2020). 


\bibitem{Agaev:2019kkz} S.~S.~Agaev, K.~Azizi, and H.~Sundu,
Nucl.\ Phys.\ B \textbf{951}, 114890 (2020).


\bibitem{Sundu:2019feu} H.~Sundu, S.~S.~Agaev, and K.~Azizi,
Eur.\ Phys.\ J.\ C \textbf{79}, 753 (2019).


\bibitem{Agaev:2019lwh} S.~S.~Agaev, K.~Azizi, B.~Barsbay, and H.~Sundu,
Phys.\ Rev.\ D \textbf{101}, 094026 (2020). 


\bibitem{Agaev:2020zag} S.~S.~Agaev, K.~Azizi, B.~Barsbay, and H.~Sundu,
Chin.\ Phys.\ C \textbf{45}, 013105 (2021).


\bibitem{Yu:2017pmn} F.~S.~Yu, 
Eur.\ Phys.\ J.\ C \textbf{82}, 641 (2022). 


\bibitem{Hughes:2017xie} C.~Hughes, E.~Eichten, and C.~T.~H.~Davies,
Phys.\ Rev.\ D \textbf{97}, 054505 (2018).


\bibitem{Esposito:2018cwh} A.~Esposito, and A.~D.~Polosa,
Eur.\ Phys.\ J.\ C \textbf{78}, 782 (2018).


\bibitem{Anwar:2017toa} M.~N.~Anwar, J.~Ferretti, F.~K.~Guo, E.~Santopinto,
and B.~S.~Zou, 
Eur.\ Phys.\ J.\ C \textbf{78}, 647 (2018).


\bibitem{Bai:2016int} Y.~Bai, S.~Lu, and J.~Osborne,
Phys.\ Lett.\ B \textbf{798}, 134930 (2019).


\bibitem{Liu:2019zuc} M.~S.~Liu, Q.~F.~L\"u, X.~H.~Zhong, and Q.~Zhao,
Phys.\ Rev.\ D \textbf{100}, 016006 (2019).


\bibitem{LHCb:2020bwg} R.~Aaij \textit{et al.} (LHCb Collaboration),
Sci.\ Bull. \textbf{65}, 1983 (2020).


\bibitem{Bouhova-Thacker:2022vnt} E.~Bouhova-Thacker (ATLAS Collaboration),
PoS \textbf{ICHEP2022}, 806 (2022).


\bibitem{CMS:2023owd} A.~Hayrapetyan, \textit{et al.} (CMS Collaboration)
arXiv:2306.07164 [hep-ex].


\bibitem{Zhang:2020xtb} J.~R.~Zhang,
Phys.\ Rev.\ D \textbf{103}, 014018 (2021).


\bibitem{Wang:2020ols} Z.~G.~Wang,
Chin.\ Phys.\ C \textbf{44}, 113106 (2020).


\bibitem{Wang:2020dlo} Z.~G.~Wang,
Int.\ J.\ Mod.\ Phys.\ A \textbf{36}, 2150014 (2021).


\bibitem{Albuquerque:2020hio} R.~M.~Albuquerque, S.~Narison,
A.~Rabemananjara, D.~Rabetiarivony, and G.~Randriamanatrika,
Phys.\ Rev.\ D \textbf{102}, 094001 (2020).


\bibitem{Yang:2020wkh} B.~C.~Yang, L.~Tang, and C.~F.~Qiao
Eur.\ Phys.\ J. C \textbf{81}, 324 (2021). 


\bibitem{Becchi:2020mjz} C.~Becchi, A.~Giachino, L.~Maiani, and
E.~Santopinto,
Phys.\ Lett.\ B \textbf{806}, 135495 (2020).


\bibitem{Becchi:2020uvq} C.~Becchi, A.~Giachino, L.~Maiani, and
E.~Santopinto, 
Phys.\ Lett.\ B \textbf{811}, 135952 (2020). 


\bibitem{Dong:2020nwy} X.~K.~Dong, V.~Baru, F.~K.~Guo, C.~Hanhart, and
A.~Nefediev,
Phys.\ Rev.\ Lett. \textbf{126}, 132001 (2021); \textbf{127}, 119901(E)
(2021). 


\bibitem{Dong:2021lkh} X.~K.~Dong, V.~Baru, F.~K.~Guo, C.~Hanhart,
A.~Nefediev, and B.~S.~Zou,
Sci.\ Bull. \textbf{66}, 2462 (2021).


\bibitem{Liang:2021fzr} Z.~R.~Liang, X.~Y.~Wu, and D.~L.~Yao,
Phys.\ Rev.\ D \textbf{104}, 034034 (2021).


\bibitem{Wang:2022xja} Z.~G.~Wang,
Nucl.\ Phys.\ B \textbf{985}, 115983 (2022). 


\bibitem{Faustov:2022mvs} R.~N.~Faustov, V.~O.~Galkin, and E.~M.~Savchenko,
Symmetry \textbf{14}, 2504 (2022). 


\bibitem{Niu:2022vqp} P.~Niu, Z.~Zhang, Q.~Wang, and M.~L.~Du,
arXiv:2212.06535.


\bibitem{Dong:2022sef} W.~C.~Dong and Z.~G.~Wang,
Phys.\ Rev.\ D \textbf{107}, 074010 (2023). 


\bibitem{Yu:2022lak} G.~L.~Yu, Z.~Y.~Li, Z.~G.~Wang, J.~Lu, and M.~Yan,
Eur.\ Phys.\ J. C \textbf{83}, 416 (2023). 


\bibitem{Kuang:2023vac} S.~Q.~Kuang, Q.~Zhou, D.~Guo, Q.~H.~Yang, and
L.~Y.~Dai,
Eur.\ Phys.\ J. C \textbf{83}, 383 (2023). 


\bibitem{Shifman:1978bx} M.~A.~Shifman, A.~I.~Vainshtein, and
V.~I.~Zakharov, 
Nucl.\ Phys.\ B \textbf{147}, 385 (1979).


\bibitem{Shifman:1978by} M.~A.~Shifman, A.~I.~Vainshtein, and
V.~I.~Zakharov, 
Nucl.\ Phys.\ B \textbf{147}, 448 (1979).


\bibitem{PDG:2022} R.~L.~Workman \textit{et al.} (Particle Data Group),
Prog.\ Theor.\ Exp.\ Phys.\ \textbf{2022}, 083C01 (2022).


\bibitem{Kiselev:2001xa} V.~V.~Kiselev, A.~K.~Likhoded, O.~N.~Pakhomova, and
V.~A.~Saleev, 
Phys.\ Rev.\ D \textbf{65}, 034013 (2002).


\bibitem{Colangelo:1992cx} P.~Colangelo, G.~Nardulli, and N.~Paver,
Z. Phys. C \textbf{57}, 43 (1993). 


\bibitem{VeliVeliev:2012cc} E.~Veli Veliev, K.~Azizi, H.~Sundu, and G.~Kaya,
PoS (Confinement X) 339, 2012; 
arXiv:1205.5703.


\bibitem{Agaev:2023gaq} S.~S.~Agaev, K.~Azizi, B.~Barsbay and H.~Sundu,
arXiv:2304.09943 [hep-ph].


\bibitem{Agaev:2023ruu} S.~S.~Agaev, K.~Azizi, B.~Barsbay and H.~Sundu,
arXiv:2305.03696 [hep-ph].


\bibitem{Maiani:2014} L.~Maiani, F.~Piccinini, A.~D.~Polosa, and V.~Riquer,
Phys.\ Rev.\ D \textbf{89}, 114010 (2014).


\bibitem{Wang:2014vha} Z.~G.~Wang,
Commun.\ Theor.\ Phys.\ \textbf{63}, 325 (2015).


\bibitem{Agaev:2017tzv} S.~S.~Agaev, K.~Azizi, and H.~Sundu,
Phys.\ Rev.\ D \textbf{96}, 034026 (2017).
\end{thebibliography}
\end{document}